\documentclass[12pt,preprint]{aastex}
\usepackage{graphicx,natbib,subfig}

\shorttitle{NIR Spectroscopy of TW Hya}
\shortauthors{Vacca \& Sandell}

\begin{document}

\title{Near-Infrared Spectroscopy of TW Hya: A Revised Spectral Type and Comparison with Magnetospheric Accretion Models}

\author{William Vacca\altaffilmark{1} and G\"oran Sandell}
\affil{SOFIA-USRA, NASA Ames Research Center, Mail Stop 211-3,
Building N211, Rm. 251}
\altaffiltext{1}{Visiting Astronomer at the Infrared Telescope Facility, which is operated by the University of Hawaii under Cooperative Agreement no. NNX-08AE38A with the National Aeronautics and Space Administration, Science Mission Directorate, Planetary Astronomy Program.}

\email{wvacca@sofia.usra.edu, gsandell@sofia.usra.edu}

\begin{abstract}

We present high signal-to-noise, moderate spectral resolution ($R \sim 2000-2500$) near-infrared ($0.8-5.0$ $\mu$m) spectroscopy of the nearby T Tauri star TW Hya. By comparing the spectrum and the equivalent widths of several atomic and molecular features with those for stars in the IRTF near-infrared library, we revise the spectral type to M2.5V, which is later than usually adopted (K7V). This implies a substantially cooler stellar temperature than previously assumed. Comparison with various pre-main sequence models suggests that TW Hya is only  $\sim 3$ Myr old; much younger than the usually adopted $8 - 10$ Myr. Analysis of the relative strengths of the H lines seen in the spectrum yields estimates for the temperature and density of the emitting region of $T_e \geq 7500$ K and $n_e \sim 10^{12} - 10^{13}$ cm$^{-3}$. The thickness of the emitting region is $10^2 - 10^4$ km and the covering fraction is $f_\ast \sim 0.04$. Our derived physical parameter values agree with the predictions of the magnetospheric accretion scenario. The highest signal-to-noise H lines have profiles that indicate multiple emission components. We derive an excess spectrum (above that of the M2.5V template) that peaks in the H band. Although our derived veiling values, $\sim 0.1$, agree with previous estimates, the excess spectrum does not match that of current models in which this flux is generated by an inner optically thin disk. We suggest that the excess flux spectrum instead reflects the differences in atmospheric opacity, gravity, and age between TW Hya and older, higher gravity field M2.5 dwarfs.
\end{abstract}

\keywords{techniques: spectroscopic - stars: pre-main-sequence - stars: individual: TW Hya}

\section{Introduction}

The optical and near-infrared (NIR) spectra of Classical T Tauri stars (CTTSs) are characterized by strong H emission lines \citep[see e.g.,][]{Gullbring98,Muzerolle98b}. In the currently accepted paradigm for these pre-main sequence objects, these emission lines arise from gas being accreted onto the stellar surface from the inner regions of a circumstellar disk 
along the stellar magnetic field lines. In this magnetospheric accretion flow model, the inflowing material is heated and ionized as it passes through a standing shock close to the stellar photosphere \citep{Calvet98}. Attempts to confirm observationally the predictions of these models have generally focussed on only a few of the strongest H lines, usually H$\alpha$, Pa$\beta$, and Br$\gamma$ \citep[e.g.][]{Muzerolle01,Folha01}. Furthermore, much of the theoretical work to date has focussed on modeling individual emission line profiles (at high resolution) and the overall spectral energy distributions (at low resolution) of CTTSs, with relatively little emphasis on the fluxes and flux ratios of a series of emission lines. Only recently have the relative strengths of the higher order members of various H series been combined with their stronger counterparts to estimate the density and temperature of the emission regions in CTTSs \citep{Herczeg04,Bary08} and thereby constrain theoretical models. A complication in carrying out such comparisons of relative line fluxes to model predictions arises from the notorious variability of the spectra of CTTSs \citep[e.g.][and references therein]{Alencar01}, and therefore the need to acquire spectra covering all useful lines simultaneously. Consequently, the location and physical conditions of the H line emitting regions are not well constrained observationally. 

At a distance of $\sim$ 51 pc \citep{Mamajek05}, TW Hya is one of the nearest T Tauri stars. Although it is a member of a small group of stars known as the TW Hydra association \citep[TWA;][]{Kastner97}, TW Hya resides surprisingly far from any known molecular cloud. \citet{Herbig78} originally identified this object as a possible post-T Tauri star with a spectral class of K7 Ve. However, its optical spectrum exhibits a very strong and variable H$\alpha$ emission line and emission lines of He I, [\ion{O}{1}], \ion{O}{1}, and \ion{Ca}{2}, features that suggest that it is a CTTS \citep{Rucinski83}.  \citet{Rucinski85} noted that TW Hya was a strong IRAS source, suggesting that it is surrounded by a dusty accretion disk. This was confirmed by \citet{Weintraub89}, who found TW Hya to be a strong sub-millimeter source. The accretion disk around TW Hya was subsequently directly imaged at various wavelengths by \citet{Wilner00},  \citet{Krist00}, \citet{Trilling01}, \citet{Weinberger02} and \citet{Qi04} and is seen approximately face-on. Analysis of the spectral energy distribution (SED) of TW Hya has indicated that the system contains a so-called  ``transitional disk" \citep{Strom89} with two separate components:  a largely dust free, optically thin, hot, inner region, extending from $\sim$ 0.06 - 4 AU \citep{Calvet02,Eisner06,Hughes07}, and a cold outer region extending to a radius of $\sim$ 200 AU. TW Hya is actively accreting material from its disk, with an accretion rate of  $\sim4 - 20 \times10^{-10}$ M$_{\odot}$~yr$^{-1}$ \citep{Muzerolle00,Alencar02,Herczeg04}. The age of the TWA has been has been estimated from a variety of methods to be between 5 and 30 Myr \citep{Kastner97,Webb99, Weintraub00b, Batalha02, Makarov05, Barrado06}. For a K7V spectral type, comparison of the location of TW Hya on the H-R diagram with pre-main sequence isochrones gives an age of TW Hya of $\sim 10$ Myr \citep{Webb99}, although a few substantially younger age estimates can be found in the literature \citep[e.g.,][]{Makarov05, Barrado06}. Summaries of the properties of TW Hya and the TWA have been given in the general reviews by \citet{Zuckerman01} and \citet{Zuckerman04}.
 
In this paper we present new high signal-to-noise medium resolution ($R\sim 2000-2500$) NIR ($0.8-5.0\mu$m) spectroscopy of TW Hya and use these data to reassess its spectral classification. We derive a new spectral type, which is substantially later than the canonical value. The later spectral type implies a much cooler stellar temperature, and hence a much younger age than usually adopted. Subtraction of a scaled, matching spectral template from our spectrum reveals numerous (primarily H) emission features. Because the spectra simultaneously cover both the Pa and Br series, we avoid the ambiguity introduced by temporal variability of the line fluxes and we use the relative line fluxes to determine the physical conditions of the emitting gas. The derived temperature, density, and thickness of the emission region are found to be in good agreement with the predictions of magnetospheric accretion models. 

\section{Observations and Data Reduction}

We observed TW Hya at the NASA Infrared Telescope Facility (IRTF) on Mauna Kea on 2010 Feb 26 and Feb 27 (UT) with SpeX, the facility near-infrared medium resolution cross-dispersed spectrograph \citep{Rayner03}. Ten individual exposure of TW Hya, each lasting 100 s, were obtained using the short-wavelength cross-dispersed (SXD) mode of SpeX on 2010 Feb 27 11:26 UT. This mode yields spectra spanning
the wavelength range 0.8 -2.4 $\mu$m divided into 6 spectral orders. Twenty exposures, each lasting 30 s, were obtained in the long wavelength cross-dispersed (LXD2.1) mode on 2010 Feb 26 11:20 UT. This mode yields spectra covering 2.2$-$5.1$\mu$m in 6 spectral orders.
The observations were acquired in ``pair mode", in which the object was observed at two separate positions along the 15\arcsec-long slit. The slit width was set to 0$\farcs$3, which yields a nominal resolving power of 2000 for the SXD spectra and 2500 for the LXD2.1 spectra. (At the distance of TW Hya of 51 pc, the SpeX $0\farcs3$ slit spans $\pm 7.7$ AU.) The slit was set to the parallactic angle during the observations. The airmass was about 1.74 for both sets of observations. Observations of HD 92845, an A0$\, $V star, used as a ``telluric standard" to correct for absorption due to the Earth's atmosphere and to flux calibrate the target spectra, were obtained immediately preceding the observations of TW Hya. The airmass difference between the observations of the object and the standard was 0.07. The seeing was estimated to be $\sim 0\farcs7-0\farcs8$ and $\sim 0\farcs6$ at 2.2 $\mu$m on 26 Feb and 27 Feb, respectively, and conditions were clear on both nights. A set of internal flat fields and arc frames were obtained immediately after the observations of TW Hya for flat fielding and wavelength calibration purposes.

\begin{figure}
\includegraphics[height=4.0in,keepaspectratio=true, viewport= 40 0 360 320 ]{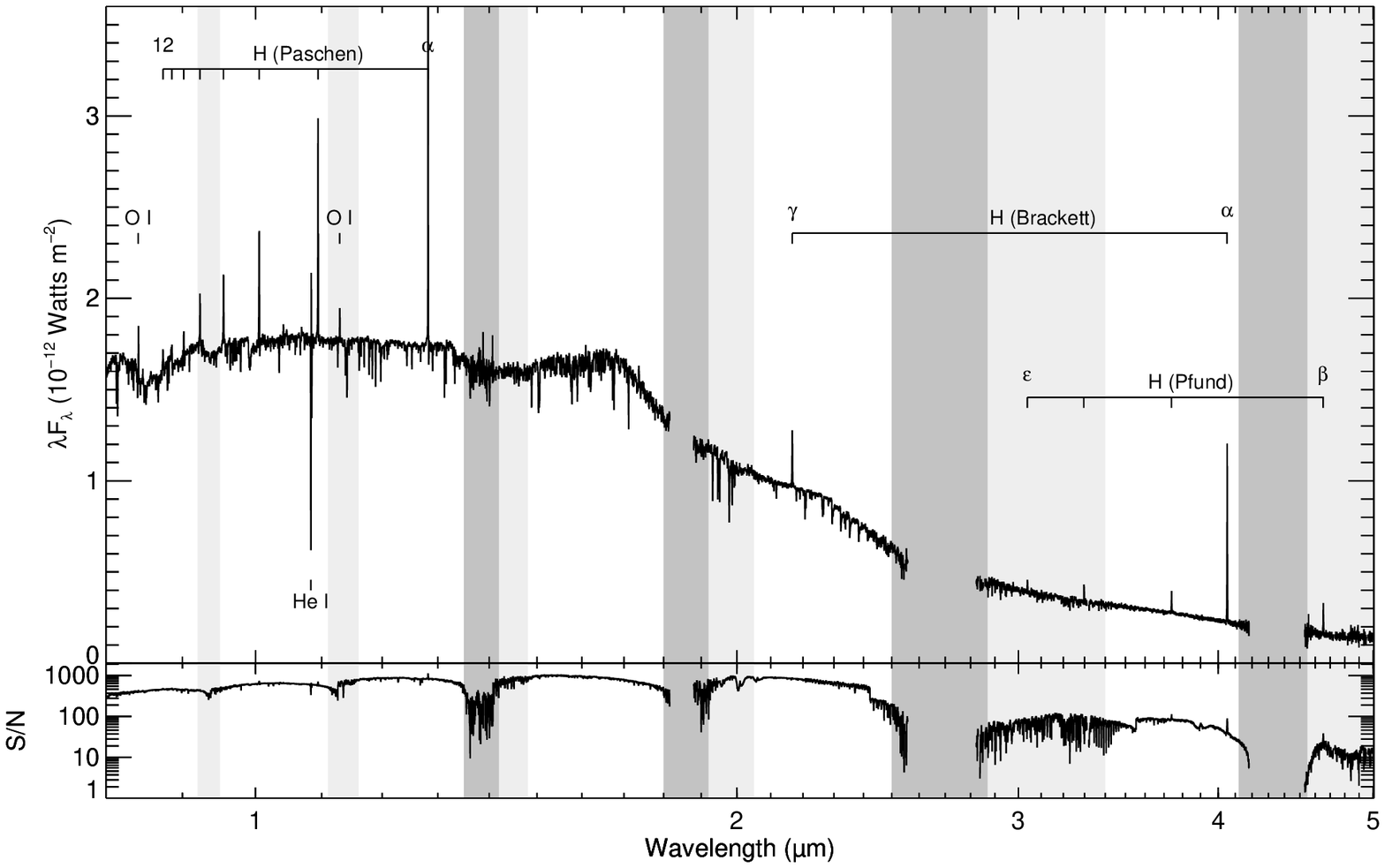}
\caption{ Flux-calibrated NIR spectrum of TW Hya obtained with SpeX on the IRTF.  
The strongest emission lines are identified. Regions of poor atmospheric transmission ($<20\%$) are shown
in dark gray, while regions of moderate atmospheric transmission ($<80\%$) are shown in light gray. This figure
can be directly compared with the figures in \citet{Rayner09}. The bottom plot shows the statistical S/N.}
\label{Fig1}
\end{figure}

The data were reduced using Spextool \citep{Cushing04}, the IDL-based package developed for the reduction of SpeX data. The Spextool package performs non-linearity corrections, flat fielding, image pair subtraction, aperture definition, optimal extraction, and wavelength calibration. The sets of spectra resulting from the individual exposures were median combined and then corrected for telluric absorption and flux calibrated using the extracted A0\, V telluric standard spectra and the technique and software described by \citet{Vacca03}. The spectra from the individual orders were then spliced together by matching the flux levels in the overlapping wavelength regions, and regions of poor atmospheric transmission were removed. The final SXD and LXD spectra were merged by scaling the SXD spectra by 1.05 in order to match the flux levels of the LXD data in the region of wavelength overlap between 2.2 and 2.4 $\mu$m and then computing a weighted average. The weights were determined from the signal-to-noise (S/N) values in the two spectra. The final $0.8-5.0$ $\mu$m spectrum  is shown in Fig.\ \ref{Fig1}. The S/N varies across the spectral range; it is of the order of several hundred across the SXD wavelength range, $\sim 50$ across the $3-4$ $\mu$m range and about $10$ in the $4-5$ $\mu$m range.

We computed synthetic NIR magnitudes in various filters from our final spectrum. The estimated 2MASS magnitudes are $J = 8.35$, $H=7.67$, and $K_s=7.41$. Comparison with the 2MASS point source catalogue indicates that our synthetic magnitudes are systematically lower by $\sim12$\%. Given the variable nature of TW Hya, it is not clear if this difference reflects a real change in the source brightness or a systematic error in our absolute flux calibration. However our synthetic $J-H$ and $H-K_s$ colors agree to within 0.02 mags with the 2MASS values, which indicates that our relative flux calibration is very accurate \citep[see also][where it is demonstrated that the adopted method of flux calibration produces spectra with relative fluxes accurate to a few percent ]{Rayner09}.

\section{Analysis}

\subsection{Spectral Classification}

Although a spectral type as early as  K6Ve has been given for TW Hya \citep{Torres06}, the most commonly quoted spectral type for this object is K7Ve following \citet{Herbig78}. However, determinations of the spectral type of T Tauri stars based on optical spectra can be biased by the substantial excess optical continuum produced by the accretion process \citep[see e.g.,][]{Mora01} as well as any differential extinction due to dust either locally or along the line of sight. In the NIR, however, the contribution from the excess continuum is believed to be small for most CTTSs, and this spectral region is much less affected by dust than the optical, while still containing numerous diagnostic lines that can be used for spectral classification. With the recent publication of the IRTF Spectral Library \citep{Rayner09,Cushing05}, it is possible to re-evaluate the spectral types for CTTSs by directly comparing their NIR spectra with those for a large stellar sample.  Therefore, we have re-examined the spectral type for TW Hya based on our SpeX spectrum, which was obtained with the same instrumentation and reduced in exactly the same manner as those in the IRTF Spectral Library. We attempted to determine the spectral classification of TW Hya using a variety of methods. 

Comparison of the overall shape $0.8-5.0$ $\mu$m spectrum of TW Hya (Fig.\ \ref{Fig1}) with those presented in the IRTF Spectral Library \citep{Rayner09} immediately suggests that this object has a spectral type between M1V and M4V, and certainly later than K7V. In particular, the broad shallow absorption at $\sim 0.85~\mu$m, due to TiO bandheads, and the prominent continuum `bumps' in the H and K bands, due to water absorption at the edges of the bands at $\sim 1.4$, $1.9$, and $2.7 \mu$m, suggest a spectral type later than about M1.5V but earlier than about M3.5V  \citep[see][]{Rayner09}. Such a comparison should not be affected by extinction along the line of sight to TW Hya because, at a distance of only 51 pc and a location well outside of any molecular cloud, as well as a nearly pole-on orientation, TW Hya is believed to have $A_V = 0$ \citep[][see \S 3.2 below]{Herczeg04}. Furthermore, the M stars in the IRTF Spectral Library either have zero extinction or their spectra have been corrected for extinction. Nevertheless, T Tauri stars are known to exhibit ``veiling", in which emission from the surrounding accretion disk provides additional flux above that of the photosphere in the IR, while the accretion shock contributes additional flux in the UV/optical to the total spectrum from the object. As TW Hya is a ``transitional disk" system, the NIR veiling is believed to be rather low \citep[$\lesssim10$\%;][see also \S 3.3 below]{JohnsKrull01}. 
Even if the veiling were substantial, the smooth excess flux spectrum expected from the disk emission should not produce the broad absorption features that characterize the spectra of early M subtypes.

We then independently examined various spectral regions, and compared them with the corresponding regions of the spectra in the IRTF Spectral Library. We carried out this comparison both ``by eye'' and using a least-squares
fitting procedure (to determine the best-fit scaling for each spectrum in the IRTF Spectral Library).  As can be seen in Figs.\ \ref{Fig2a} and \ref{Fig2b}, both the spectral slope and the depths of various spectral features in the $0.8-1.35$ $\mu$m region indicate that TW Hya has a spectral type between M1V and M3V. In particular, the presence of weak TiO bands at $\sim 0.85$ and $0.89$ $\mu$m, a weak absorption due to FeH at $0.87$ $\mu$m, and the absorption feature due to H$_2$O at 1.33 $\mu$m suggest a spectral type of early M, rather than K. A similar result is found for the K-band region (Fig.\ \ref{Fig2d}), in which the spectral shape of TW Hya appears to be very similar to that of an M2.5V star. For the H-band, however, we find a slight disagreement between the spectral shape, which indicates a spectral type of M2V or later, and the strengths of various discrete features (particularly the complex of Fe and Si lines between 1.60 and 1.65 $\mu$m), which indicate a spectral type of $\sim$ M1V. The $\chi^2$ comparison over the wavelength region between 0.95 and 1.34 $\mu$m (where veiling is expected to be minimal; see below) yields low $\chi^2$ values for the M0.5V, M1.5V, M2.5V, and M3V spectra from the Library, with a formal best match with the M1.5V spectrum. However, the overall spectrum does not match that of the M0.5V star and the shape of the K band spectral region does not match that of the M1.5V star. There is very little difference between the M2.5V and M3V spectra.

We also measured the equivalent widths (EWs) of numerous features in our spectrum and compared them with values determined from direct measurements of the spectra of K and M stars available in the IRTF Spectral Library \citep{Rayner09,Cushing05}. (Since our spectrum has the same resolution and comparable S/N as those in the IRTF Spectral Library, a comparison with these values should provide a better indication of the spectral type than those given by other authors with different stellar libraries.) The EWs of some of these absorption features (due to \ion{Al}{1}, \ion{K}{1}, \ion{Mg}{1}, \ion{Na}{1}, and FeH) in the spectrum of TW Hya are presented in Table \ref{abslines}. A comparison between a subset of these values deemed to be the most sensitive diagnostics of spectral type and those derived in an identical manner from the K and M stars in IRTF Spectral Library is shown in Figure \ref{Fig3} as a function of spectral type. In this figure, the horizontal dashed lines denote the range in the EW  ($\pm 1 \sigma$) for TW Hya. The results presented in Fig. \ref{Fig3}, particularly the FeH and \ion{K}{1} $1.253 \mu$m EWs, clearly suggest a spectral type of M2.5V - M3.5V for TW Hya. 

Because of veiling, one must be careful in interpreting the EWs as spectral type diagnostics: lines may appear to be too weak simply because of the additional continuum flux from the disk. Therefore, the effect of veiling on a spectral type determination depends on the trend of the EW of a feature with spectral type for normal stars. If the EW of a feature decreases with spectral type, veiling will cause a feature to appear weaker and will result in a later spectral classification. If the EW increases with spectral type, veiling will result in an earlier spectral classification. However, veiling cannot produce lines that are too {\it strong} for a given spectral type. Furthermore, the veiling in TW Hya is expected to be negligible in the Y and J bands \citep{Edwards06, Salyk09}. For these reasons, we relied primarily on the diagnostic lines in these bands and on those lines whose strength increases with later spectral type. The latter set of lines should yield a limit on the spectral type. Nevertheless, even for the H and K bands, the effect of veiling on the spectral type we determine from the EWs should be small. The true EW (${\rm EW}_0$) can be computed from the observed EW (${\rm EW}_{\rm obs}$) with 
\begin{equation}
{\rm EW}_0 = {\rm EW}_{\rm obs} (1 + r)
\end{equation}
where $r$ is the veiling value. \citet{JohnsKrull01} give a K band veiling value for TW Hya of $r_K = 0.07 \pm 0.04$.
Even a correction of 10\% to the observed EW values will not yield a spectral type substantially different from what we determine from the observed (uncorrected) EWs. Furthermore, such an increase in the EWs will tend to yield later, not earlier, spectral types from the most sensitive lines.

The effects of veiling can be mitigated to a large extent by considering pairs of closely spaced diagnostic lines sensitive to spectral class. The ratio of the EWs of two such lines is given by 
\begin{equation}
\frac{{\rm EW}_0^{(1)}}{{\rm EW}_0^{(2)}} = \frac{{\rm EW}_{obs}^{(1)}}{{\rm EW}_{obs}^{(2)}} \cdot \frac{1 + r_1}{1 + r_2}~~~.
\end{equation}
We can express the veiling of one line in terms of the other
\begin{equation}
r_1 = r_2 + \Delta r
\end{equation} 
and if the lines are suitably close in wavelength then $\Delta r << r_2$. Therefore,
\begin{equation} 
 \frac{{\rm EW}_0^{(1)}}{{\rm EW}_0^{(2)}} = \frac{{\rm EW}_{obs}^{(1)}}{{\rm EW}_{obs}^{(2)}} \cdot \Bigl(1 + \frac{\Delta r}{1 + r_2}\Bigr) \approx \frac{{\rm EW}_{obs}^{(1)}}{{\rm EW}_{obs}^{(2)}}
 \end{equation}
and the observed EW ratios will reflect the true EWs. If this ratio is sensitive to spectral type, it can be used to classify the star. Fig.\ \ref{EWratio} demonstrates the use
of this technique for TW Hya using the \ion{Na}{1} 1.14 $\mu$m and \ion{Mg}{1} 1.18$\mu$m lines. This figure demonstrates again that the spectral type for TW Hya is approximately M2.5V, independent of any possible veiling. The agreement between this determination and that derived from the individual lines demonstrates that the veiling around 1 $\mu$m must be negligible in TW Hya.

Based on the results of these three comparison methods, we conclude that TW Hya has a spectral type of M2.5V or M3V, with a possible range between M1.5 V and M3.5 V. For this spectral type, the calibrations for M stars given by \citet{Leggett96} and \citet{ReidHawley} yield an effective temperature of $3400\pm 200$ K and a mass of $0.4\pm 0.1 M_\odot$. To estimate the stellar radius, we adopted the 2MASS J band magnitude of 8.22, a distance of 51 pc, and the J-band bolometric corrections from the \citet{Baraffe98} models to compute the bolometric luminosity. The radius was then determined from 
\begin{equation}
R_\ast = \Bigl(\frac{L_{bol}}{4 \pi \sigma T_{eff}^4}\Bigr)^{0.5}~~~.
\end{equation}
With this procedure, we find $L_{\rm bol} = 0.19\pm 0.03 L_\odot$ and $R_\ast = 1.3\pm 0.2 R_\odot$. The derived luminosity is in good agreement with other estimates \citep[$0.22-0.23 L_\odot$;][]{Kastner97,Webb99}.

\subsection{Excess Flux - Emission Line Spectrum}

As can be seen in Fig.\ 1, the NIR spectrum of TW Hya consists of numerous emission lines atop a photospheric spectrum. The emission lines presumably arise from the accretion process. We generated the spectrum of this emission component by scaling the spectrum of an M2.5V star (Gl 381) from the IRTF Spectral Library to that of TW Hya and subtracting it. The scale factor was determined to be 0.29 via a least squares calculation between 0.95 and 1.34 $\mu$m. This wavelength region was chosen for two reasons: (1) as stated above, previous investigations have found that the veiling in the Y band (i.e., around 1 $\mu$m) is negligible \citep{Edwards06, Salyk09}; (2) matching a library spectrum at any longer or shorter wavelengths resulted in substantial regions with negative `excess' fluxes and veiling factors. Since the extinction toward TW Hya is negligible, the latter result is not physically justified. The result of the subtraction of the scaled M2.5V template spectrum is shown at the bottoms of Figs.\ \ref{Fig2a}-\ref{Fig2d}. 
Numerous lines of the hydrogen Paschen (Pa), Brackett (Br), Pfund (Pf), and Humphreys (Hu) series, as well as emission lines of \ion{O}{1}  and \ion{Ca}{2}  appear in the subtracted spectrum.  We see no evidence of the H$_2$ emission at 2.1218 $\mu$m reported by \citet{Weintraub00}. The measured line fluxes are presented in Table \ref{emisslines}. We now discuss the properties of the H, He, and O lines.

\subsubsection{\rm \ion{H}{1}}

As shown by \citet{Bary08} for a sample of 15 classical T Tauri stars and by \citet{Najita10} for the case of TW Hya itself, the relative H line strengths in the spectra of these objects can be well reproduced with the Case B recombination approximation. Although the relative line strengths within any given series can be used to estimate the temperature and density of the emitting region, a more sensitive diagnostic is provided by the inter-series ratios of line strengths \citep[e.g.,][]{Bary08}. Therefore, we computed the flux of each line relative to that of Pa $\beta$, the strongest line in our spectrum. 
We then used the theoretical H line fluxes for Case B as provided by \citet{Storey95} and \citet{Hummer87} to compute the corresponding line ratios for electron temperatures between 500 and 30000 K and electron densities between $10^2$ and $10^{14}$ cm$^{-3}$. For each temperature and density pair we computed a $\chi^2$ value between the theoretical and observed line ratios. Contours of the $\chi^2$ values are shown in Fig.\ \ref{chicont}. The minimum $\chi^2$ was found for $T_e = 20000$K and $n_e = 10^{13}$ cm$^{-3}$ and the corresponding theoretical line ratios for these values are plotted along with the observed line flux ratios as a function of upper level quantum number in Fig \ref{lineratio}.  However, as seen in Fig.\ \ref{chicont}, the electron temperature is not tightly constrained by the line ratios and we find that any temperature $T_e \geq 7500$ K can provide an adequate representation of the data points. For temperatures $7500 \leq T_e \leq 12500$ K, the best fit electron density is $n_e = 10^{12}$ cm$^{-3}$, while for temperatures $T_e \geq 12500$ K, the best fit yields $n_e = 10^{13}$ cm$^{-3}$. In our analysis below we will use both $n_e = 10^{12}$ cm$^{-3}$ and $n_e = 10^{13}$ cm$^{-3}$, with the corresponding temperature values. For comparison, we also plot in Fig \ref{lineratio} the theoretical values for $T_e= 3000$ K and $n_e = 10^{10}$ cm$^{-3}$ derived by \citet{Bary08}. This comparison clearly demonstrates that the values found by \citet{Bary08} cannot reproduce the observed line ratios for TW Hya. We also analyzed the H line strengths seen in the {\it Spitzer} IRS spectra  of TW Hya presented by \citet{Najita10} in the same manner. Because of the large error bars and relatively few data points, the H flux ratios relative to Hu$\alpha$ in the SH2 spectrum provide only weak constraints on the electron temperature and density. Although the formal best fit for the SH2 line fluxes is found for $T_e= 30000$ K and $n_e = 10^{9}$ cm$^{-3}$, there is very little difference between the theoretical line flux ratios for this set of parameters and the values we found from our NIR line analysis. For the SH1 spectrum of \citet{Najita10}, the best fit was found for $T_e= 5000$ K and $n_e = 10^{10}$ cm$^{-3}$, close to the values advocated by \citet{Bary08}.

As stated above, it is the combination of multiple H line series that provide the strongest constraints on the temperature and density. With just the Pa H series alone, for example, it would be nearly impossible to distinguish lines produced in a region with $n_e = 10^{4}$ cm$^{-3}$ from those produced in a region with $n_e = 10^{13}$ cm$^{-3}$. However, the fluxes of the H lines in the spectra of T Tauri stars are known to vary, and our LXD data were not acquired simultaneously with the SXD data (the time difference was almost exactly 24 hours). Therefore, we also carried out our analysis using only the Pa and Br lines in our SXD 0.8-2.5 $\mu$m spectrum. We obtained results similar to those given above: a temperature of $3 \times 10^4$ K and a density $\sim 10^{14}$ cm$^{-3}$ provided the best fit in this case, but $T_e = 20000$ K, and $n_e = 10^{13}$ cm$^{-3}$ were nearly equally acceptable. This, along with the fact that the observed flux levels of the SXD and LXD spectra differed by only 5\% between the two nights, suggests that TW Hya did not vary substantially between the two observations, and confirms the results found by including the Pf lines in the analysis. We note that the H line strengths in the spectrum of TW Hya obtained by \citet{Covey10} in 2008 are only about 15-20\% larger than the values we measure.

The measured H lines contain several sets for which the $N_{upper}$ is the same (e.g., Pa $\beta$ and Br $\alpha$, or Pf $\beta$ and Br $\gamma$ and Pa $\delta$). The flux ratios of these lines are fairly insensitive to the density and temperature of the emitting gas, particularly at densities above $10^9$ cm$^{-3}$, and hence can be used to estimate the extinction.
The agreement with the Case B ratios shown in Fig.\ \ref{lineratio} indicates that the assumption of zero reddening, and hence low dust content in the emitting region, is correct. 


We used the flux in the Pa $\beta$ and Br $\gamma$ lines to estimate the accretion rate using the calibrations of \citet{Muzerolle98b}. These lines yield accretion luminosities of $L_{\rm acc} = 5.9 \times 10^{-2} L_\odot$ (from Pa $\beta$) and $2.9 \times 10^{-2} L_\odot$ (from Br $\gamma$), and mass accretion rates of $6.1$ and $2.9 \times 10^{-9} M_\odot {\rm yr^{-1}}$, respectively, from the relation
\begin{equation}
\dot{M} = \frac{L_{\rm acc} R_\ast}{G M_\ast}
\end{equation}
which can be expressed as 
\begin{equation}
\log \dot{M} ~[{\rm M_\odot ~ yr^{-1}}] = \log \Bigl(\frac{L_{\rm acc}}{L_\odot}\Bigr) + \log \Bigl(\frac{R_\ast}{R_\odot}\Bigr) - \log \Bigl(\frac{M_\ast}{M_\odot}\Bigr) - 7.50 ~~~.
\end{equation}
These values are considerably higher than that estimated by \citet{Muzerolle00} for TW Hya, but are well within the range found by \citet{Najita07} for transitional disk systems in Taurus. Furthermore, the accretion luminosities are well within the ranges of values determined by \citet{Batalha02} from their analysis of optical data for TW Hya. Under the assumption that the Pa $\beta$ line is optically thin, the line luminosity also provides an estimate of the ionization rate from
\begin{equation}
Q_0 = \frac{L_{{\rm Pa} \beta} \alpha_B}{\epsilon_{{\rm Pa}\beta}}
\end{equation} 
or
\begin{equation}
Q_0 = 3.8 \times 10^{43} \Bigl(\frac{L_{{\rm Pa} \beta}}{L_\odot}\Bigr) \Bigl(\frac{\alpha_{\rm B}}{10^{-13}}\Bigr) \Bigl(\frac{\epsilon_{\rm Pa \beta}}{10^{-26}}\Bigr)^{-1}\quad {\rm ph ~s^{-1}} ~~~.
\end{equation}
We find the ionization rate to be $Q_0 \sim 4.5-6.6 \times 10^{42}$ ph s$^{-1}$ for $n_e = 10^{12}$ cm$^{-3}$ and $Q_0 \sim 6.1-7.0 \times 10^{42}$ ph s$^{-1}$ for $n_e = 10^{13}$ cm$^{-3}$. The values derived for a density of $10^{12}$ cm$^{-3}$ are in good agreement with that derived by \citet{Najita10}, while the values found for $n_e = 10^{13}$ cm$^{-3}$ are about a factor of two larger. 

The volume of the ionized region is given by 
\begin{equation}
V = \frac{Q_0}{n_e^2 \alpha_B} = \frac{L_{{\rm Pa} \beta}}{\epsilon_{{\rm Pa}\beta} n_e^2}
\end{equation}
or 
\begin{equation}
V = 3.8 \times 10^{35}  \Bigl( \frac{L_{{\rm Pa} \beta}}{L_\odot}\Bigr) \Bigl(\frac{\epsilon_{\rm Pa \beta}}{10^{-26}}\Bigr)^{-1} \Bigl(\frac{n_e}{10^{12}}\Bigr)^{-2}\quad {\rm cm^{3}}\quad .
\label{veqn}
\end{equation}
where we have assumed $n_p \approx n_e$. The emitting volume was found to be surprisingly small, $V \sim 0.6-4.0 \times 10^{31}$ cm$^{3}$ for $n_e = 10^{12}$ cm$^{-3}$ and $1.4-3.3 \times 10^{29}$ cm$^{3}$ for $n_e = 10^{13}$ cm$^{-3}$. We derive a mass of the emitting gas, $m_H n_p V$,  of between $1.1-6.2 \times 10^{19}$ g (for $n_e = 10^{12}$ cm$^{-3}$) and $2.3-5.6 \times 10^{18}$ g (for $n_e = 10^{13}$ cm$^{-3}$). These values are substantially lower (factors of $10^4-10^5$) than that derived by \citet{Najita10}, who assumed a density of $n_e < 10^{8}$ cm$^{-3}$.
Assuming the emitting volume is located at or near the surface of the star, we can estimate the thickness of the emitting region from 
\begin{equation}
\Delta R = \frac{V}{4 \pi R_\ast^2 f_\ast}~~.
\label{dR}
\end{equation}
where $R_\ast$ is the radius of the star and $f_\ast$ is the fraction of the surface area covered by the accretion spot. Combining this with Eqn.\ \ref{veqn}, we have
\begin{equation}
\Delta R = 6.3 \times 10^{9} \Bigl(\frac{L_{{\rm Pa} \beta}}{L_\odot}\Bigr) \Bigl(\frac{R_\ast}{R_\odot}\Bigr)^{-2} \Bigl(\frac{\epsilon_{\rm Pa \beta}}{10^{-26}}\Bigr)^{-1} \Bigl(\frac{n_e}{10^{12}}\Bigr)^{-2} \Bigl(\frac{f_\ast}{0.01}\Bigr)^{-1} \quad {\rm km}~~~.
\label{dRscale}
\end{equation}
If we require that the optical depth in the Pa $\beta$ line $\tau_{\rm Pa \beta} < 1$, as suggested by the good fit of the theoretical Case B values to the observed line ratios, we can place a lower limit on $f_\ast$. Using the line opacity factors, $\Omega$, given by \citet{Storey95}, we have
\begin{equation}
f_\ast > \frac{\Omega_{\rm Pa \beta}}{\epsilon_{\rm Pa \beta}} \frac{L_{\rm Pa \beta}}{4 \pi R_\ast^2} ~~~
\end{equation}
and we find $f_\ast > 0.01$ for $n_e = 10^{12}$ cm$^{-3}$ and $f_\ast > 0.02$ for $n_e = 10^{13}$ cm$^{-3}$. These limits are in reasonable agreement with the values of $f_\ast$ derived by \citet{Batalha02}. Using their relation between $f_\ast$ and $L_{\rm acc}$
\begin{equation}
f_\ast = 0.49 (L_{\rm acc}/L_\odot) + 0.01~~~,
\end{equation}
we derive $f_\ast = 0.04$ from the Pa $\beta$ luminosity. 
Using this value in Eqns.\ \ref{dR} and \ref{dRscale}, we derive the thicknesses corresponding to the emission volumes of $1.6 - 9.1 \times 10^4$ km for $n_e = 10^{12}$ cm$^{-3}$ and $3.4 - 8.2 \times10^2$ km for $n_e = 10^{13}$ cm$^{-3}$. The ranges of thicknesses in turn imply column densities of $N_H \sim 1.6 - 9.1 \times 10^{21}$ and $3.4 - 8.2 \times 10^{20}$ cm$^{-2}$, respectively. Since TW Hya is seen nearly pole-on, the relatively small volumes and thicknesses we derive suggests that the H lines arise primarily from a fairly thin region, consisting of one or more hot spots located presumably at the base of the accretion column at the surface of the star near the magnetic pole.

In Fig.\ 
\ref{Hlineprof} we show the profiles of the Pa $\beta$ and Br $\gamma$ emission lines. Overplotted is a Gaussian with a FWHM of 150 km s${-1}$. (Recall that our spectra have $R=2000$ in this wavelength range, which corresponds to a velocity resolution of $\sim 150$ km s$^{-1}$.) The high S/N afforded by our spectra allows us to discern multiple components in these profiles, which are replicated in many of the stronger H lines: a narrow (unresolved) core at the rest wavelength, a weaker component that extends over a fairly large velocity range (from roughly $-350$ to $+250$ km s$^{-1}$), and an inverse P Cygni absorption feature on the red side at $\sim 300$ km s$^{-1}$. 
We note that extended wings are not seen in the (substantially weaker) \ion{O}{1} and \ion{Ca}{2} line profiles, whereas they are present on even the weak, higher order Paschen lines, a result that suggests that they are not due to our point spread function. In addition, the Pa $\delta$ line was used to generate the telluric correction \citep[see][]{Vacca03} and so should be the least affected by any residuals from the telluric correction process, yet it clearly exhibits broad wings. Also, the broad emission wings are not seen on many of the H lines in the reduced spectrum of TW Hya, but appear only after subtraction of the M2.5V template. Furthermore, broad emission over a range of velocities up to $-400$ km s$^{-1}$ is not uncommon in CTTSs \citep[e.g][]{Edwards06}.
The H line profiles suggest that the strongest emission arises at the base of the accretion column (zero velocity), with additional emission from an outflowing wind with velocities up to about $300$ km s$^{-1}$, and absorption from gas that is falling onto the star with a velocity of $\sim 300$ km s$^{-1}$.

\subsubsection{\rm \ion{He}{1}}

Exhibiting two absorption troughs on either side of an emission peak at the line center, the He I line profile (Fig.\ \ref{Helineprof}) is more complicated than those of the H lines. The blue shifted absorption component has a minimum at $\sim -100$ km s$^{-1}$ and extends to about $-300$ km s$^{-1}$, while the weaker redshifted component has a minimum at $\sim 300$ km s$^{-1}$ and extends to about $+500$ km s$^{-1}$. The velocity of the red absorption feature is the same as that seen in the H line profiles, which suggests that the H and He absorption features are formed in the same region of the accretion flow or infall onto the star. Our  \ion{He}{1} 1.083 $\mu$m line profile is unlike any of those presented by \citet{Edwards03}, who observed the \ion{He}{1} 1.083 $\mu$m line in a sample of six YSOs. Nor is it similar to the \ion{He}{1} 1.083 $\mu$m line profile presented by \citet{Dupree05} from their observations of TW Hya in 2002. In the \citet{Edwards06} sample of 39 CTTSs (which includes TW Hya), the only profile the current observations of \ion{He}{1} 1.083 $\mu$m resembles is that of CY Tau, although the extent in velocity of the blueshifted absorption we observe (with our limited velocity resolution of $\sim 150$ km s$^{-1}$) agrees with that of the blueshifted emission seen in the TW Hya spectrum presented by \citet{Edwards06}. These comparisons indicate that the He I line profile in TW Hya is remarkably variable. This conclusion is strengthened by the profile of \ion{He}{1} 1.083 $\mu$m in the recently published spectrum of TW Hya obtained by \citet{Covey10} in 2008 using SpeX at the IRTF. The \ion{He}{1} 1.083 $\mu$m profile in their spectrum exhibits a considerably stronger emission peak, a somewhat shallower P Cygni absorption trough, and almost no evidence of the red absorption feature. This change in the \ion{He}{1} line profile suggests a denser outflow during the observations of \citet{Covey10}, which would be consistent with their somewhat stronger H lines.

\subsubsection{\rm \ion{O}{1}}

As discussed by \citet{Grandi75} and summarized by \citet{Rudy89}, the permitted \ion{O}{1} lines seen in the spectrum of TW Hya can arise from recombination, fluorescence excited by the stellar continuum, or resonance fluorescence excited by Ly $\beta$. Because continuum fluorescence should generate an 0.8446 $\mu$m line that is considerably stronger than the 1.1287 $\mu$m line, while the observations indicate that these two lines have approximately equal strength, this excitation mechanism can be ruled out. The relative line strengths produced by recombination can be derived from the volume emission coefficients,
\begin{equation}
j_{8446} = n_e n_{\rm O II} \frac{h \nu_{8446}}{4 \pi} \alpha_{8446}^{eff}~~~~,
\end{equation}
\begin{equation}
j_{11287} = n_e n_{\rm O II} \frac{h \nu_{11287}}{4 \pi} \alpha_{11287}^{eff}~~~~,
\end{equation}
and
\begin{equation}
j_{\rm Pa \beta} = n_e n_{\rm H II} \frac{h \nu_{\rm Pa \beta}}{4 \pi} \alpha_{\rm Pa \beta}^{eff}~~~.
\end{equation}
Because \ion{O}{1} and \ion{H}{1} have very similar ionization potentials, we might expect O and H to have similar spatial distributions and ionization fractions in the emission regions. In this case, the strength of the 0.8446 $\mu$m line relative to Pa $\beta$ can be expressed as 
\begin{equation}
\frac{j_{8446}}{j_{\rm Pa \beta}} = \Bigl(\frac{n_{\rm O}}{n_{\rm H}}\Bigr) \frac{\lambda_{\rm Pa \beta}\alpha_{8446}^{eff}}{\lambda_{8446}\alpha_{\rm Pa \beta}^{eff}}~~~~.
\end{equation}
This is provides an upper limit to the flux ratio because it assumes all the O is in the form of  \ion{O}{2}; some O may be in \ion{O}{3} and higher ionization stages.

Extrapolating the effective recombination coefficients of \citet{EscVic92} for \ion{O}{1} to electron temperatures of $\sim 10^4$ K, and using the H line emissivities given by \citet{Storey95} and a solar oxygen abundance \citep{Asplund10}, we estimate line strength ratios resulting from recombination of $j_{8446}/j_{11287} \approx 2 $ and $j_{8446}/j_{\rm Pa \beta} \lesssim 2.6\times 10^{-3}$. The observed line ratios are found to be $F_{8446}/F_{11287} \approx 1.4$ and  $F_{8446}/F_{\rm Pa \beta} \approx 7.7 \times 10^{-2}$. Although it could be argued that the former ratio could be affected by reddening, which could be responsible for reducing the observed ratio below the expected value, the latter ratio is already $\sim 30$ times stronger than predicted and would only increase with any corrections for reddening. Furthermore, to bring the former line ratio into agreement with the predictions would require a reddening correction corresponding to $A_V \approx 1.9$, which seems far too large for TW Hya. Therefore, under the assumption that the H and \ion{O}{1} emission lines are generated in the same region, we can rule out recombination as the source of the \ion{O}{1} emission. This leaves fluorescence by Ly $\beta$ as the excitation mechanism for the \ion{O}{1} lines. This mechanism should produce equal numbers of photons in the 0.8446 and 1.1287 $\mu$m lines, which is exactly what we observe. The fact that we do observe equal numbers of photons in the lines again confirms that there is no differential reddening along the line of sight to the emitting region and very little dust within the emission region itself, as found previously by many others \citep[e.g.][]{Muzerolle00}.

\subsection{Excess Flux - Continuum Spectrum}

In Fig.\ \ref{veilplot} we present the $0.8-5~\mu$m spectrum of TW Hya with the scaled IRTF Library spectra of a K7V star (HD 237903) and an M2.5V star (Gl 381) overplotted. The middle panels show the result of subtracting the scaled library spectra from the observed spectrum, while the bottom panels present the ratios of this excess spectrum to the scaled library spectral templates. 
Both library spectra were scaled to match the observed spectrum of TW Hya between 0.95 and 1.34 $\mu$m. Template spectral types of M2V - M3.5V yield very similar excess emission spectra to that shown in the middle right panel of Fig.\ \ref{veilplot}. As seen in the middle left of Fig.\ \ref{veilplot}, earlier template spectral types yield excess emission spectra with several broad ``features",  particularly between the J, H, and K photometric bands, and somewhat stronger excess emission longwards of $\sim 3$ $\mu$m, but wholly unphysical negative values shortward of $\sim 1.1 \mu$m. 

The excess flux spectrum we derive from subtracting the M2.5V star exhibits a rise to shorter wavelengths ($\lambda < 1 ~\mu$m) that is most likely due to the (possibly optically thick) continuum emission produced by the accretion process (``optical veiling''). The excess at longer wavelengths could be due to the thermal emission from the accretion disk, either the optically thick disk at $R>4$ AU \citep{Calvet02} or an optically thin inner disk \citep{Eisner06}. However, the excess spectrum cannot be fit with any simple disk model; the model of \citet{Eisner06} is shown as a dotted line in this panel and clearly does not match the shape of the derived spectrum, greatly underestimating the flux in the H and K bands and overestimating the flux at longer wavelengths. The peak in the excess flux spectrum in the H band requires a dust temperature of $\sim 1600$ K, but optically thin dust disk models, with power law distributions for the temperature and surface density, generally cannot reproduce the sharp drop off in flux on either side of this peak that is seen in our data. 

An `excess flux' spectrum that peaks in the H band is very reminiscent of the spectral difference observed between late type giants and dwarfs. As can be seen in the series of spectra presented in the IRTF Spectral Library \citep{Rayner09}, for a given spectral type a late type giant star exhibits a prominent H band excess over that of a dwarf star due to the decrease in the $\rm H^-$ opacity at the lower density and gravity \citep{Wing03, Mihalas}. To demonstrate this, in Fig.\ \ref{giant_dwarf} we plot the `excess' spectrum derived by scaling the spectrum of an M2.5V star to that of an M2.5III star, both taken from the IRTF Spectral Library, over the wavelength range $0.9-1.34 \mu$m, and subtracting the former from the latter. The resulting difference spectrum was then scaled to match the level of the H band excess in the spectrum of TW Hya and overplotted on the TW Hya excess spectrum. As can be seen from this figure, the shapes of the excesses are nearly identical, with the giant $-$ dwarf excess extending from the H band through the K band and into the M band. Clearly, of course, TW Hya is not an M giant. Nevertheless, the mass and radius values determined above yield a gravity of $\log g = 3.8$, far below the value of $\sim 4.8$ estimated for field M2.5 dwarfs from the parameters given by \citet{Leggett96} and \citet{ReidHawley}. Therefore, we believe the TW Hya H and K band excess is in fact not indicative of disk emission, but rather is a signature of TW Hya's youth (relative to field M dwarfs), as it contracts and evolves down to the main sequence. This conjecture might be tested with higher resolution observations of atomic absorption lines, whose widths would reflect the value of $\log g$, in much the same manner as described by \citet{Cruz09} for L dwarfs. If the excess could be calibrated as a function of $\log g$, it could provide a separate estimate of the age of the system.

Under the assumption that the excess spectrum is due to emission arising from the accretion process, the ratio of the excess spectrum to the library template yields the veiling $r$. As seen in the bottom right panel of Fig.\ \ref{veilplot},  the veiling derived from the M2.5V spectral template 
is fairly small: $\sim 6$\% between 0.8 and 0.9 $\mu$m,  $\sim 8$\% between 1.4 and 1.8 $\mu$m, $\sim 10$\% between 2.0 and 2.4 $\mu$m, and $\sim 12$\% between 3.0 and 4.0 $\mu$m. 
Our derived veiling values are in good agreement with the estimate of \citet{JohnsKrull01} for the K band and those of \citet{Alencar02} for $0.8-0.9~\mu$m. However, if the suggestion given above for the shape of the excess spectrum is correct, only a fraction of this excess emission arises from the accretion process and therefore our veiling estimates for the H, K and L bands are strict upper limits. Although it is uncertain how much of the excess longwards of $\sim1$ $\mu$m is due to the effects of the lower gravity of TW Hya relative to the comparison field M dwarf used for the subtraction, it is possible that this difference in gravity is responsible for nearly all of the observed excess in the H and K bands, and a significant fraction of the excess in the L band.
 
\section{Discussion}
\subsection{Spectral Type}
The wavelength region around 1 $\mu$m is ideal for spectral classification of CTTSs: (1) the excess emission arising from the accretion process, which is very strong in the UV and optical wavelengths, is relatively small; (2) the emission from the accretion disk itself, which can dominate the spectrum at K band and longer wavelengths, is minimal; (3) the effects of any extinction are mitigated substantially compared to the optical; and (4) there are numerous strong photospheric absorption features present that can be used for classification. The last point is particularly important because, while the contributions from the accretion shock and circumstellar disk (UV and IR excesses, respectively) are at a relative minimum in the Y and J bands, they are not necessarily negligible and evidence for significant J band veiling has been found in some sources \citep{Folha99, Cieza05}. However, the ratio of absorption EWs, as discussed in \S 3.1, provides a veiling independent method of determining the spectral type of a source. Although the relatively `clean' nature of this spectral region has been pointed out before \citep{Kenyon90} and exploited to study the emission associated with the accretion process in CTTSs \citep{Edwards03, Edwards06}, this is the first time of which we are aware that the absorption features in this region have been used for spectral classification of a CTTS. The absorption line EWs we measure, in comparison with those derived from the IRTF library spectra, provide strong evidence that the underlying star in TW Hya has a spectral type of $\sim$ M2.5V. This result suggests that the values for the stellar mass and radius typically adopted in the literature \citep[$T_{\rm eff} \sim 4000$ K, $M \sim 0.6 M_\odot$,  and $R \sim 1 R_\odot$; e.g.,][]{Webb99,Muzerolle00,Eisner06} are incorrect, as they are based on an incorrect spectral type (K7V). We suggest that the appropriate values for TW Hya are  $T_{\rm eff} \sim 3400$ K, $M \sim 0.4 M_\odot$, and $R \sim 1.3 R_\odot$, based on an M2.5V spectral type, the temperature and mass calibrations given by \citet{Leggett96} and \citet{ReidHawley}, and the estimated bolometric luminosity.

The spectral type we determine, and the stellar parameters we derive, are clearly inconsistent with those found by others \citep[e.g.][]{Webb99,Yang05}. Examination of the spectra shown by \citet{Webb99} indicates that the K7V spectral classification is appropriate, as TW Hya clearly has an earlier type than those of the M stars shown. \citet{Alencar02} and \citet{Yang05} fit atmosphere models to high resolution optical spectra and derived parameters again consistent with a K7V star. On the other hand, the mass estimated for TW Hya from the \citet{Yang05} stellar parameters is far too large for even a K7V spectral type. Furthermore, the moderate resolution optical spectrum presented by \citet{Rucinski83} (their Figure 9), with its strong rise in the continuum to the red and its sharp and prominent TiO bandheads near 7100 \AA, is more suggestive of an early M dwarf than a late K dwarf (see below) and does not seem to be completely consistent with the spectrum presented by \citet{Webb99}. Can these two discrepant spectral type classifications be reconciled? The optical spectrum of TW Hya suffers from considerable and variable veiling \citep[see e.g.,][]{Alencar02}, so it is possible that veiling has contributed to mis-classification, especially if the spectral comparison is restricted to very small wavelength regions lacking strong lines diagnostic of spectral type. To investigate this possibility, we selected an M2V star from the Indo-US Spectral Library \citep{Valdes04} and added veiling in an attempt to reproduce a K7V star spectrum between 6400 and  7600 \AA\  (the spectral region displayed in Fig.\ 1b of \citet{Webb99}). We found that it was not possible unless the veiling was substantially larger ($r \sim 3$) than that estimated by either \citet{Alencar02} or \citet{Yang05} in this wavelength range. Therefore, veiling cannot be completely responsible for the classification differences. We also considered the possibility that TW Hya is an unresolved binary. However, given the relative luminosities of K and M stars, it is not physically possible for a coeval K7V+M2.5V binary system to generate a spectrum in which the K7V component dominates the optical spectrum but is nearly invisible in the NIR spectrum. 

Because both the continuum and line emission are know to vary substantially in TW Hya, it is tempting to consider the possibility that the spectrum (and spectral type) of the object varies as a function of both accretion rate and wavelength. A physically plausible explanation for this variation is that the spectrum of TW Hya consists of two separate components, one due to the star (M2.5V) and one arising from an accretion hot spot, either on the surface of the star or located in the inner disk, or from the inner disk itself. In this scenario, the hot spot or inner disk would have a temperature of  $\sim 4000$ K and would generate a spectrum similar to that of a K7V star. This component would also vary in strength with the accretion rate, dominating the (optical) spectrum and completely overwhelming the underlying M2.5V stellar spectrum when the accretion rate is high. Support for this suggestion can be found by comparing the observed optical spectra with the reported H$\alpha$ equivalent widths, a proxy for the accretion rate. The optical spectrum presented by \citet{Webb99} is clearly that of a K7V star and yields an ${\rm EW (H\alpha)} = -220$ \AA. Similarly, the optical spectrum shown by \citet{Yang05} is consistent with that of a K7V star and has  ${\rm EW (H\alpha)} \sim -146$ \AA\ (C.\ M.\ Johns-Krull, private communication). The optical spectrum presented by \citet{Rucinski83}, however, corresponds to spectral type intermediate between a K7V and an early M star, and has ${\rm EW (H\alpha)}$ between  $\sim -71$ and  $-101$ \AA . In fact, when we attempted to reproduce the features seen in this spectrum we found that a combination of a K7V and an M2.5V, with roughly equal contributions at 0.8 $\mu$m, provides a much better representation than any single K or early M spectrum alone. The shape and depth of the TiO bands at 7100 \AA\ and the simultaneous presence of these bands and a strong \ion{Ca}{1} line at 7148 \AA\  indicates that the observed spectrum is a composite of both a K7V and an early M type spectrum. In our NIR spectrum of TW Hya, the region shortward of 1$\mu$m can also be well fit with an M2.5V spectrum and a small contribution from a K7V spectrum, comprising $\sim 30$\% of the total flux at 0.8 $\mu$m. This is shown in Fig.\ \ref{M+K}, where it can be seen that at wavelengths shorter than $\sim 0.6\mu$m we would expect that, during our observations of TW Hya, the K7V spectrum contributed roughly half of the total flux, but comprised only a very small fraction of the flux longward of $1 \mu$m. Unfortunately a more satisfactory test of this suggestion cannot be carried out with the spectra and accretion rates or EW values found in the literature. (Surprisingly few optical and NIR spectra of TW Hya are actually shown in the voluminous literature on this object.)  Frequent simultaneous monitoring of TW Hya in both the optical and NIR wavelength regimes would be able to confirm or rule out our suggestion.

\subsection{Age}
Our results for the spectral type and effective temperature require a re-evaluation of the age of TW Hya. The age adopted for TW Hya under the assumption of a K7V spectral type with $T_{\rm eff} = 4000$ K is often $\sim 10$ Myr, derived by comparing the location of TW Hya in the HR diagram with model isochrones \citep[e.g.][]{Webb99}. In Fig.\ \ref{HRD}, we plot the location of TW Hya on the H-R diagram, using our estimated $T_{\rm eff}$, along with the model tracks and isochrones of \citet{Baraffe98}, \citet{D'Antona98}, \citet{Siess00}, and \citet{Tognelli10}. Despite the considerable variation in the model predictions, all of the models yield masses smaller than our adopted value of $0.4 M_\odot$ and ages substantially younger than the canonical value of 10 Myr. Large underestimates of both mass and age by the various models have been pointed out previously by \citep{Hillenbrand08}. However, even accounting for an underestimate of the age by a factor perhaps as large as two \citep{Hillenbrand08} would be insufficient to make the star 10 Myr old. The models of \citet{Baraffe98} provide a mass of $0.35 < M < 0.4 M_\odot$, closest to our adopted value of $0.4 M_\odot$. For these models, the age of TW Hya estimated from the isochrones is $\sim 3$ Myr. We note that the new location of TW Hya on the H-R diagram actually brings this source into better agreement with other members of the TWA, as can be seen from Fig.\ 3 of \citet{Webb99}, which indicates that, aside from TWA 6 and TWA 9 A and B, the locations of the members of the TWA on the H-R diagram are consistent with an isochrone age for the association of $\sim 3-5$ Myr. This age is also in reasonable agreement with the expansion age of the TWA of $4.7 \pm 0.6$ Myr derived by \citet{Makarov05}, and therefore alleviates to a considerable degree the previous discrepancy between the isochrone age of TWA and dynamical age estimates. Furthermore, compared to the usually adopted age of 10 Myr, an age of 3-5 Myr for TW Hya is in much better agreement with estimates of the half life of accretion disks in low mass stellar systems of $< 3$ Myr \citep{Evans09}. 
By 10 Myr the disks in such systems have usually transformed into debris disks with masses of $\sim 0.01-0.1 M_\earth$ \citep{Wyatt08}. The large mass of the disk in TW Hya \citep[$0.06 M_\odot$][]{Calvet02} therefore suggests an age of considerably less than 10 Myr \citep{Wyatt08}, in agreement with our much younger estimate.
We also note that TW Hya exhibits many of the same emission features, and has exactly the same Li 6708\AA\ absorption EW, as the M2.5 star GN Tau \citep{White03}. The latter star is a low-mass T Tauri star with an estimated age of $\sim 3$ Myr, again providing support for our age estimate for TW Hya.

\subsection{Accretion Parameters}
The determination of an accurate spectral type permits the subtraction of the correct spectral template to reveal the emission from the accreting gas and the circumstellar disk. Under the assumption that the H lines are optically thin and follow the Case B predictions, our analysis of the H line flux ratios yields a temperature of $\sim 20000$ K, a density of $10^{13}$ cm$^{-3}$ and a thickness of the emission column of $\Delta R \sim 10^3$ km. Comparison of our results with those presented by \citet{Calvet98}, who modeled the emission from the accreting column of gas in the magnetospheric infall scenario, reveals surprisingly good agreement for some of the physical parameters. The models of \citet{Calvet98} include emission from the shock and post-shock region, the pre-shock region, and the heated stellar atmosphere. The emission properties of the models are parametrized in terms of the energy flux carried by the accretion column into the shock, $\mathcal{F}$, given by 
\begin{equation}
\mathcal{F} = 9.8 \times 10^{10} \Bigl(\frac{\dot{M}}{10^{-8} M_\odot {\rm yr^{-1}}}\Bigr) \Bigl(\frac{M}{0.5 M_\odot}\Bigr) \Bigl(\frac{R}{2 R_\odot}\Bigr)^{-3} \Bigl(\frac{f_\ast}{0.01}\Bigr)^{-1} ~~~ {\rm erg ~cm^{-2} s^{-1}}.
\end{equation}
The H line emission in these models arises from the pre-shock region, which is predicted to have a density given by 
\begin{equation}
n_H  = 5.8 \times 10^{12} \Bigl(\frac{\dot{M}}{10^{-8} M_\odot {\rm yr^{-1}}}\Bigr) \Bigl(\frac{M}{0.5 M_\odot}\Bigr)^{-0.5} \Bigl(\frac{R}{2 R_\odot}\Bigr)^{-1.5} \Bigl(\frac{f_\ast}{0.01}\Bigr)^{-1} ~~~{\rm cm^{-3}}.
\end{equation}
For the stellar parameters for TW Hya given above, these equations give
$\mathcal{F} \sim 4.4 \times 10^{10}$ erg  cm$^{-2}$ s$^{-1}$ and $n_H \sim 1.9 \times 10^{12}$ cm$^{-3}$. The latter value is fairly close to our empirical results (under the assumption that all of this material becomes fully ionized in the pre-shock region). The models also predict a temperature for this region of $T \sim 20000$ K, and thickness of the fully ionized zone of $\Delta R \sim 10^3  -10^4$ km, again in good agreement with our findings. Recently \citet{Eisner10} has used interferometry to determine the stellocentric radii of the Br $\gamma$ emitting gas in T Tauri and Herbig Ae/Be stars. They find that the emission in general arises from gas within 0.01 AU ($1.5 \times 10^6$ km), an upper limit that our results certainly satisfy. The infall velocity in the models of \citet{Calvet98} is predicted to be 
\begin{equation}
v = 275~ \Bigl(\frac{M}{0.5 M_\odot}\Bigr)^{0.5} \Bigl(\frac{R}{2 R_\odot}\Bigr)^{-0.5} ~~~~{\rm km ~s^{-1}}
\end{equation}
or $\sim 300 ~ {\rm km ~s^{-1}}$ for the mass and radius we derive for TW Hya, in excellent agreement with the velocities corresponding to the observed inverse P Cygni absorption features seen in our H and \ion{He}{1} 1.083$\mu$m line profiles.

Using our derived density and placing a limit on the optical depth of the Pa $\beta$ line, we derive a lower limit to the fractional area covered by the accretion column of $f_\ast \gtrsim 0.02$. Our limit is well within the range of previous estimates of the covering fraction, for both TW Hya itself as well as other CTTSs \citep{Herbst88,Batalha02, Calvet98}. Although our limit is a factor of 7 larger than that estimated by \citet{Muzerolle00}, we also derive substantially larger (factors of 4-5) values of the accretion luminosity and rate than these authors. (\citet{Muzerolle00} also derive a substantially larger value of the accretion energy flux $\mathcal{F}$ than we do, largely because of this difference in the values of the covering fractions $f_\ast$.) If the covering fraction is correlated with the accretion rate, as claimed by \citet{Batalha02}, then the considerably larger accretion rates we measure compared to \citet{Muzerolle00} should be associated with a larger covering fraction. It is interesting to note that the values of the energy flux and covering fraction we derive for TW Hya are very similar to those found by \citet{Calvet98} for CY Tau, the star whose \ion{He}{1} 1.083 $\mu$m line profile resembles that in our spectrum of TW Hya most closely.

Despite the good agreement between our temperature and density estimates and the model predictions of \citet{Calvet98}, our values do not agree with those of \citet{Bary08} for their sample of CTTSs, and our analysis rules out their values for TW Hya with high confidence. \citet{Bary08} do not present results for individual stars, so it is unclear how much variation they find from object to object. The mass accretion rates listed for the objects in their sample are substantially higher than that measured for TW Hya. (The minimum value listed in their Table is more than double the value derived for TW Hya in \S 3.2.1 from the Pa $\beta$ line and five times larger than that derived from the Br $\gamma$ line.) However, one would expect that higher mass loss rates would generate larger volume densities. At the moment, we have no satisfactory explanation for this discrepancy.

The high densities we derive provide a reasonable explanation for why other lines typical of ionized regions (e.g., forbidden lines of \ion{O}{2} and \ion{O}{3}) are weak or absent in optical and NIR spectra of TW Hya obtained with narrow slits centered on the star \citep[e.g.,][]{Rucinski83}. Since these ionization states are found primarily within the region where H is fully ionized and the density is well above the critical density for these transitions, the ground-state levels giving rise to these transitions are collisionally de-excited. Spectra obtained with wide slits, encompassing lower density regions farther from the star, might be expected to reveal the usual forbidden lines \citep[see e.g.,][]{Najita10}.

Our analysis suggests that the \ion{O}{1} lines seen in the NIR spectrum are produced by Ly $\beta$ fluorescence. It is not immediately obvious, however, where the \ion{O}{1} emission is generated. \citet{Muzerolle98a} demonstrated that the profile of the 0.8446 line correlates with that of Pa 11 in a small sample of CTTSs, and therefore suggested that the lines were formed in a common region. Furthermore, the \ion{O}{1} lines in our spectrum of TW Hya are unresolved at our resolution, which indicates that they are not arising from a region experiencing substantial infall or outflow, and therefore it is tempting to suggest that they are formed at the base of the accretion column, along with the H lines. However, production of the \ion{O}{1} lines by Ly $\beta$ fluorescence requires a high Ly $\beta$ flux density in a region that contains neutral oxygen. The latter requirement is inconsistent with completely ionized H gas in the pre-shock region in the models of \citet{Calvet98}. A partially ionized region at the outer edge of the pre-shock region could be responsible for the \ion{O}{1} emission. This region could also give rise to the [\ion{O}{1}] 6300, 6363 lines that are seen in the optical spectrum \citep{Alencar02,Herczeg07}. However, fluorescence also requires a large optical depth in the Balmer transitions. Using the equation given by \citet{Grandi80} for the escape probability of H$\alpha$, $\epsilon_{\rm H\alpha}$ and that by \citet{Capriotti65} for the optical depth $\tau_{\rm H\alpha}$ in terms of $\epsilon_{\rm H\alpha}$, we derived a rough estimate of the H$\alpha$ optical depth of $\tau_{H\alpha} \sim 84$. Such a large optical depth in H$\alpha$ seems incompatible with the finding that the observed H line strength ratios are well reproduced by the theoretical Case B values \citep[see also][]{Bary08}, which assumes that all transitions above the Ly series are optically thin. A more detailed model of the emission line region and its spectrum will be needed to resolve this puzzle.

\subsection{Excess Spectrum}
Finally, our derived NIR excess flux spectrum, usually interpreted as arising from the hot inner accretion disk, does not match that from any recent model of TW Hya. While the H-band peak in our excess flux spectrum suggests a dust temperature of $\sim 1600$K, which would be expected for dust in the inner disk, the sharp drops in flux at both shorter and longer wavelengths are not reproduced by current optically thin disk models. In fact, as discussed above, we suspect that most of this excess emission does not arise from the inner accretion disk, but is rather due to the lower H$^-$ opacity in the star's atmosphere relative to that of an older, higher gravity field dwarf. Therefore, during our observations, any NIR emission from the hot inner accretion disk must have been very small, much smaller than either previous estimates or the predictions of current models. It is possible that previous values of the NIR excess have been overestimated as a result of both the adoption of an incorrect spectral type for TW Hya, as well as the spectral differences due to differences in the opacity, gravity, and age between TW Hya and any comparison field dwarf. Matching a library template with a spectral type earlier than M2V to the short wavelength end of the NIR spectrum would yield substantially more excess emission at wavelengths $> 2$ $\mu$m than we estimate. 
In addition, scaling earlier spectral type templates to match the observed TW Hya flux levels between 0.95 and 1.34 $\mu$m yields an excess spectrum for which synthetic photometry is in reasonable agreement with some of the models \citep[e.g.][]{Eisner06}. Nevertheless, in these cases the resulting excess flux spectra appear to be unphysical (because of negative values), and are certainly not the smooth blackbody functions characterized by a single dust temperature expected from the models or observed in other CTTS sources \citep[e.g.][]{Muzerolle03}.  We caution that any attempts to determine the NIR excess emission arising from accretion disks around young stellar objects similar to TW Hya must account for any possible differences in age and gravity between the target objects and the spectral templates. Failure to do so could lead to spurious results. As we have shown here, the use of an old, higher gravity field dwarf as a spectral template can lead to an excess spectrum that primarily reflects differences in the H$^-$ atmospheric opacity between the CTTS and the template, not the emission from an accretion disk.

 \section{Conclusions}
 
 We have presented new high signal-to-noise, moderate resolution NIR spectra of TW Hya obtained with SpeX at the IRTF. These data have enabled us to revise the spectral classification of this object. We find that the spectrum is more consistent with a spectral type of M2.5V, rather than the commonly adopted K7V. Based on this spectral classification, we derive new values for the effective temperature ($3400$ K), mass ($0.4 M_\odot$), radius ($1.29 R_\odot$), and age ($3$ Myr) of the system.  Subtraction of an M2.5V template drawn from the IRTF Cool Star Spectral Library reveals emission lines of H, He, Ca, and O.  The H lines consist of a narrow component, a much weaker broad component, and a red absorption feature. Analysis of the strengths of the narrow H lines indicates that they arise from a hot ($T_e \sim 20000$ K), extremely dense ($n_e \sim 10^{13}$ cm$^{-3}$) region with a filling factor of $f_\ast > 0.02$ and a thickness of $\Delta R \sim 10^2-10^3$ km. The temperature and thickness values agree remarkably well with the predictions from the model of magnetospheric accretion developed by \cite{Calvet98}. We interpret the structure in the line profile as a reflection of the relative locations of the emitting regions: the narrow core arises from the base of the accretion shock column at the stellar photosphere, while the inverse P Cygni absorption arises from the infall of gas in the accretion column and the extended wings arise from the outflow of material in a disk or coronal wind. Analysis of the \ion{O}{1} permitted lines indicates that they are generated by Ly $\beta$ fluorescence. The location of the mostly neutral region generating the \ion{O}{1} lines, however, is uncertain. We also find very little excess continuum emission arising from the optically thin hot inner accretion disk. The spectrum of the excess emission we derive does not appear to agree with the model of the inner disk presented by \citet{Eisner06}. Rather the shape of the excess suggests that it results from the lower H$^-$ opacity in the atmosphere of TW Hya, due to the lower $\log g$ and younger age, compared to the older, higher $\log g$ value for field M2.5V stars used for the spectral comparison.

\acknowledgments

WDV thanks M.\ C.\ Cushing for providing software useful for some of the analysis and plots presented in this paper, as well as for discussions about M stars. We also thank J.\ Najita for discussions regarding TW Hya, C.\ Johns-Krull for providing H $\alpha$ equivalent width measurements, and C.\ Howard for comments on the manuscript.

\newpage

\begin{subfigures}

\begin{figure}
\centering
\includegraphics[height=7in]{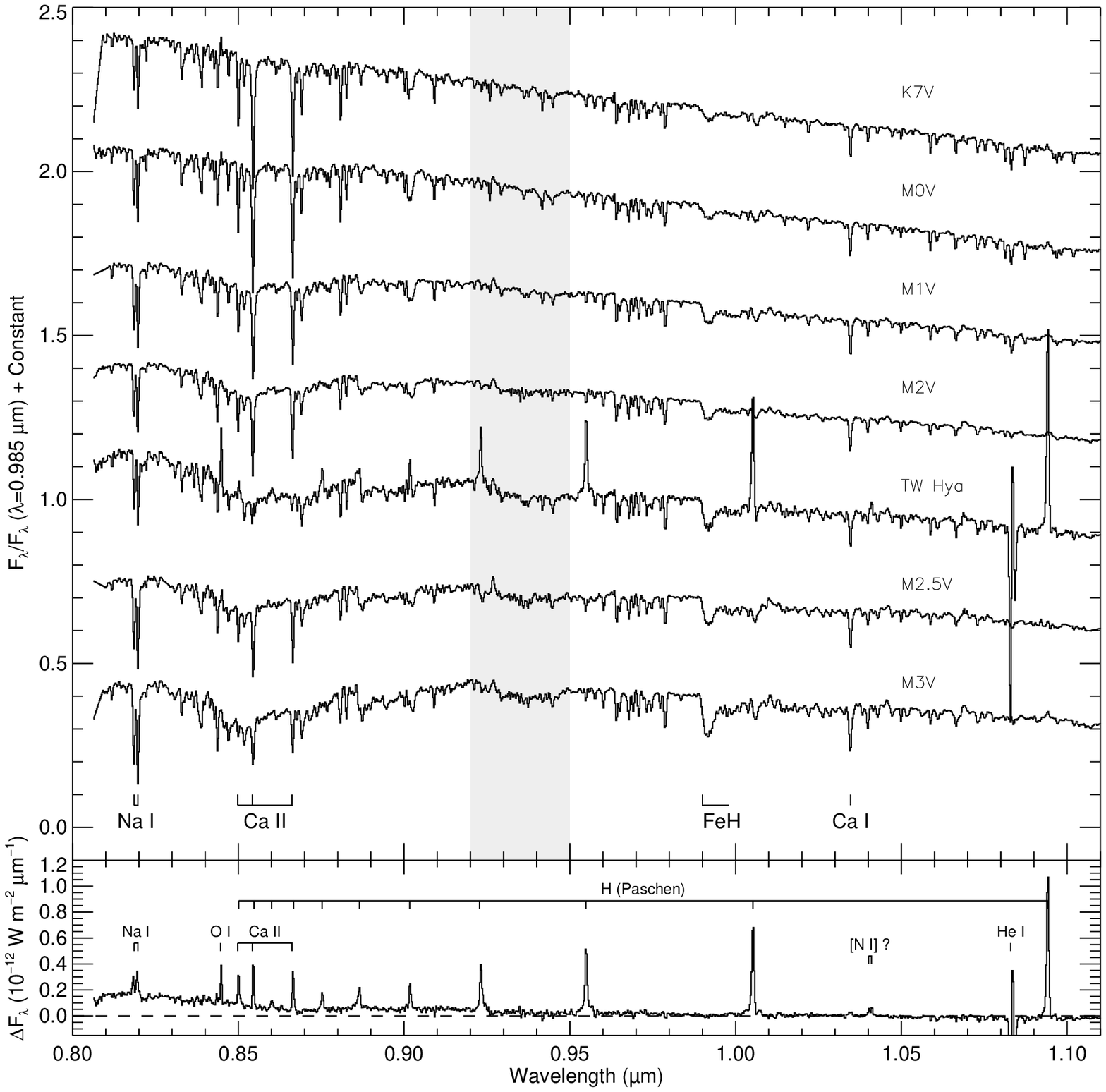}
\caption{(Top) Comparison of the normalized $0.8-1.1$ $\mu$m spectrum of TW Hya with those of late K and early M dwarfs
stars in the IRTF Library. Note the variation in the strength of the FeH absorption with spectral type. The identification
of other features can be found in \citet{Rayner09,Cushing05}. (Bottom) Residual spectrum obtained by subtracting the scaled M2.5V spectrum from that of TW Hya. }
\label{Fig2a}
\end{figure}

\begin{figure}
\centering
\includegraphics[height=7in]{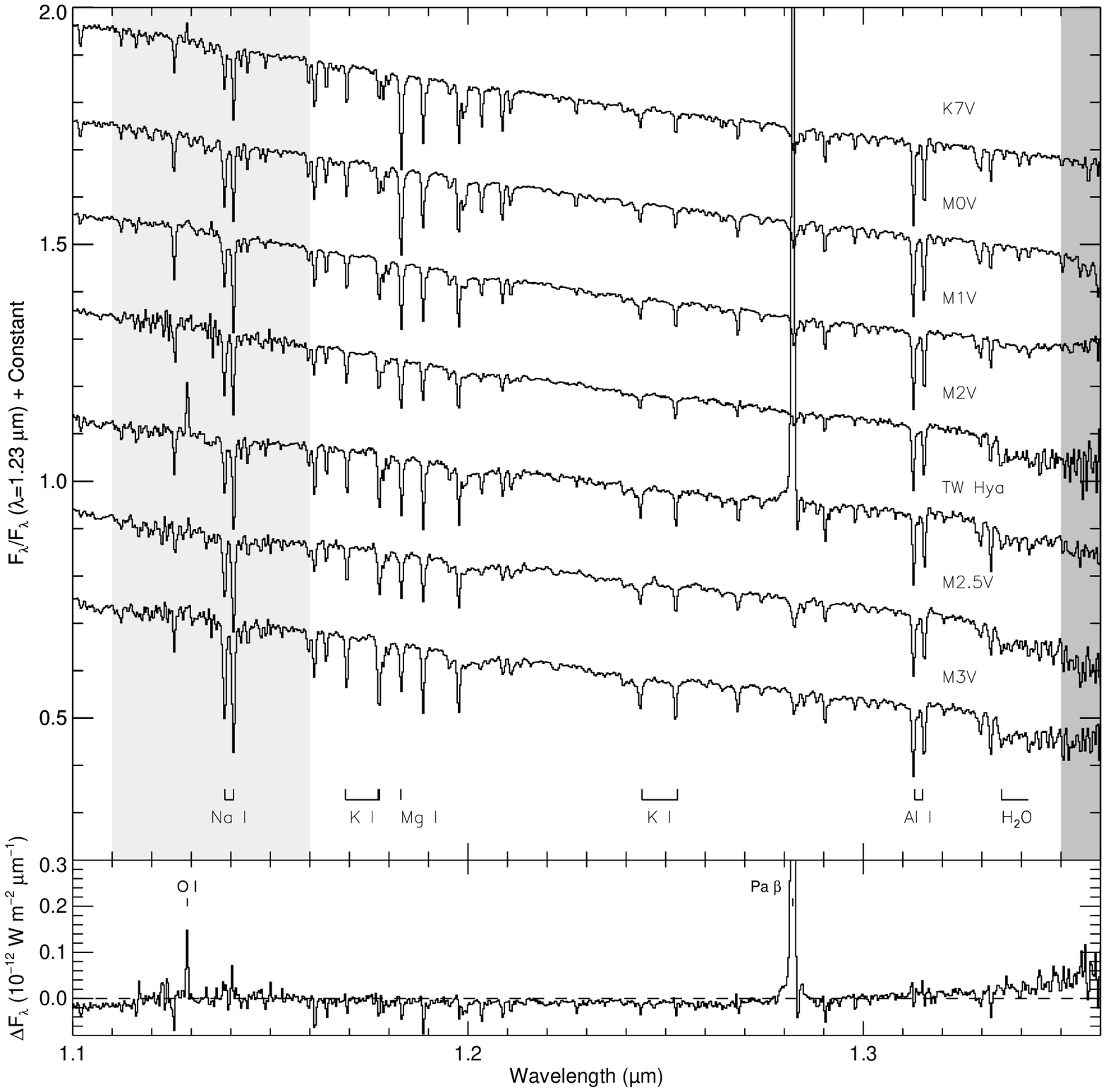}
\caption{(Top) Comparison of the normalized $1.1-1.35$ $\mu$m spectrum of TW Hya with those of late K and early M dwarfs
stars in the IRTF Library. Note the variation in the strength of the K I and H$_2$O absorption features with spectral type. 
The identification of other features can be found in \citet{Rayner09,Cushing05}. (Bottom) Residual spectrum obtained by subtracting the scaled M2.5V spectrum from that of TW Hya. }
\label{Fig2b}
\end{figure}

\begin{figure}
\centering
\includegraphics[height=7in]{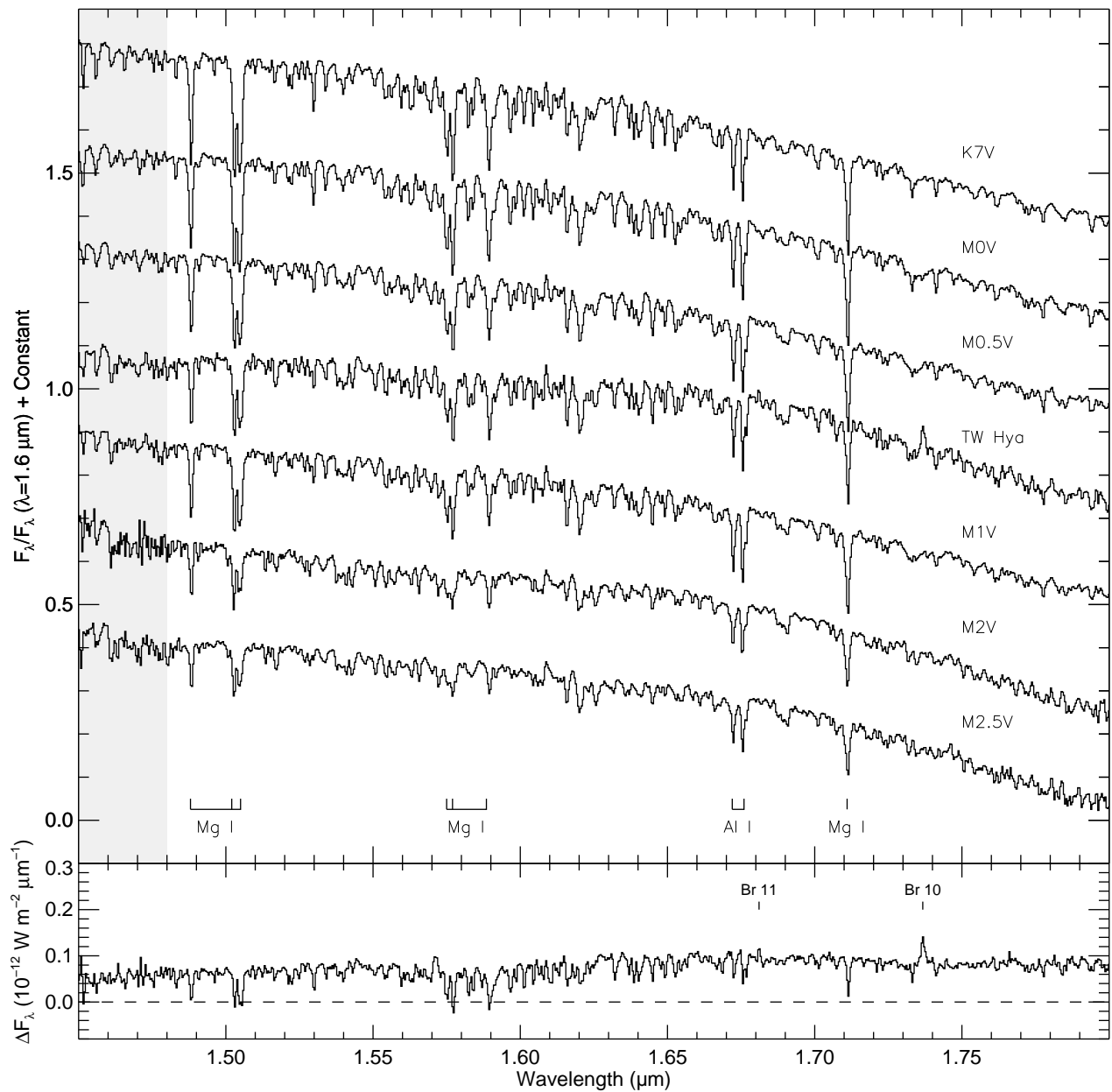}
\caption{ Comparison of the normalized $1.4-1.8$ $\mu$m spectrum of TW Hya with those of late K and early M dwarfs
stars in the IRTF Library. Note the variation in the strength of the Mg and Al absorption features, as well as the complex of Fe and Si
lines between $1.60$ and $1.65$ $\mu$m, with spectral type. 
The identification of other features can be found in \citet{Rayner09}. (Bottom) Residual spectrum obtained by subtracting the scaled M2.5V spectrum from that of TW Hya. }
\label{Fig2c}
\end{figure}

\begin{figure}
\centering
\includegraphics[height=7in]{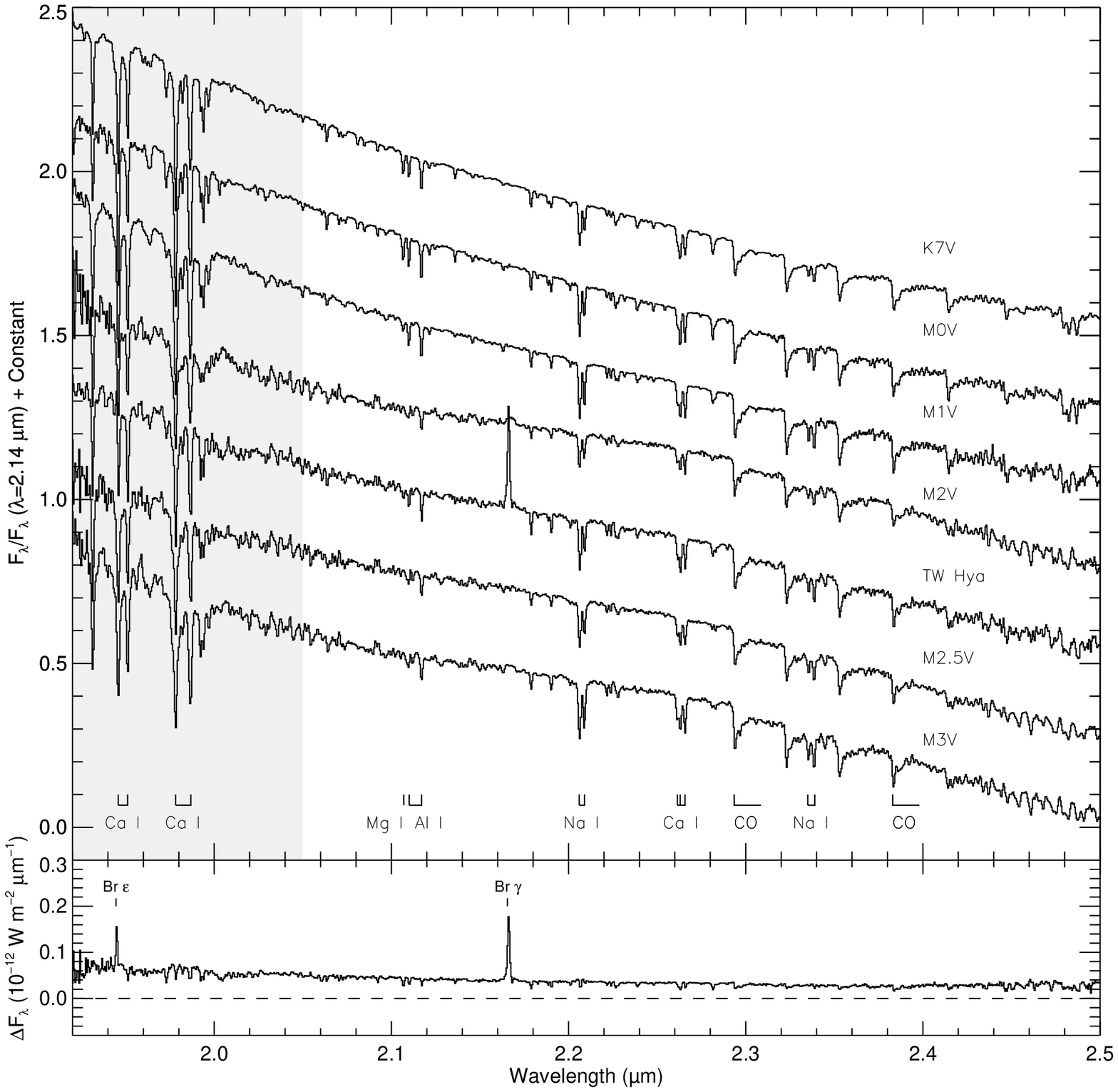}
\caption{ Comparison of the normalized $1.95-2.5$ $\mu$m spectrum of TW Hya with those of late K and early M dwarfs
stars in the IRTF Library. Note the variation in the strength of the Mg absorption feature with spectral type. 
The identification of other features can be found in \citet{Rayner09,Cushing05}. (Bottom) Residual spectrum obtained by subtracting the scaled M2.5V spectrum from that of TW Hya. }
\label{Fig2d}
\end{figure}

\end{subfigures}

\begin{figure}
\centering
\includegraphics[height=7in,keepaspectratio=true]{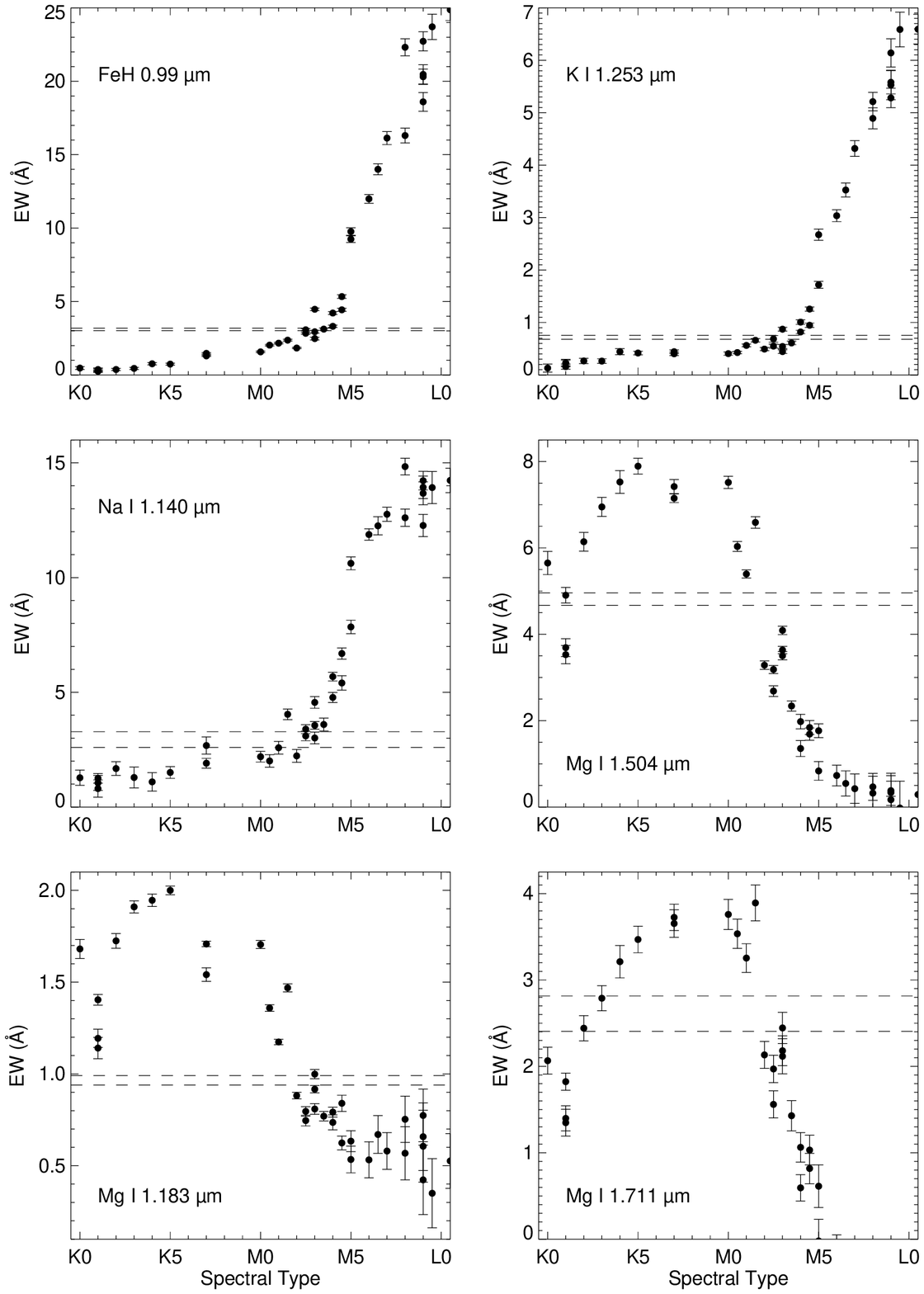}
\caption{ The strength of EW of various diagnostic spectral features as a function of spectral type for K-M dwarfs in the IRTF Spectral Library. The dashed lines denote the $\pm 1 \sigma$ range in the EWs for TW Hya. }
\label{Fig3}
\end{figure}

\begin{figure}
\includegraphics[height=5in, keepaspectratio=true]{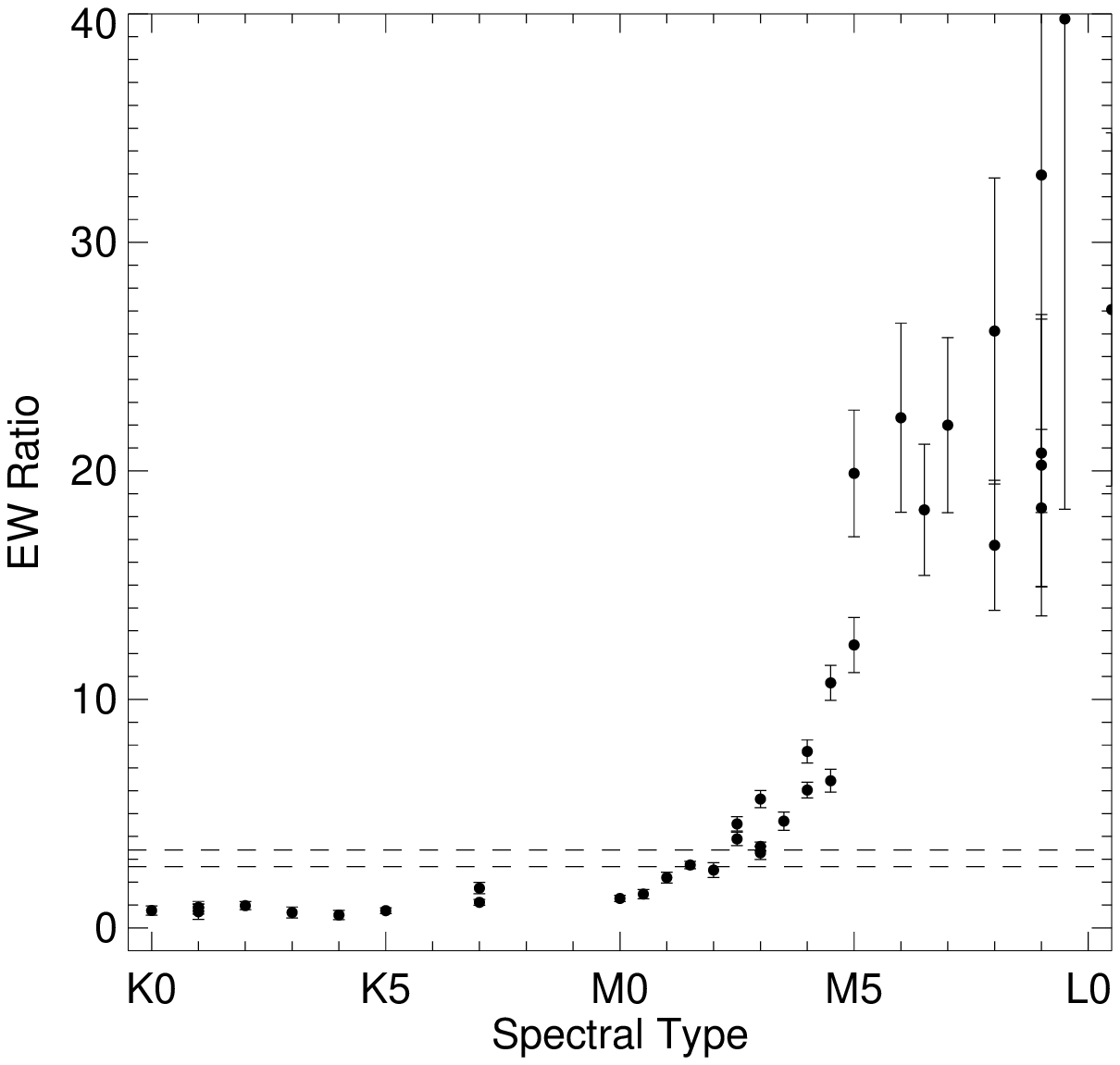}
\caption{Ratio of the EW of the \ion{Na}{1} 1.140 $\mu$m feature to the EW of the \ion{Mg}{1} 1.183 $\mu$m feature for K-M dwarfs from the IRTF spectral library. The dashed lines denote the $\pm 1 \sigma$ range in the EW ratio for TW Hya. }
\label{EWratio}
\end{figure}

\begin{figure}
\includegraphics[height=5in, keepaspectratio=true]{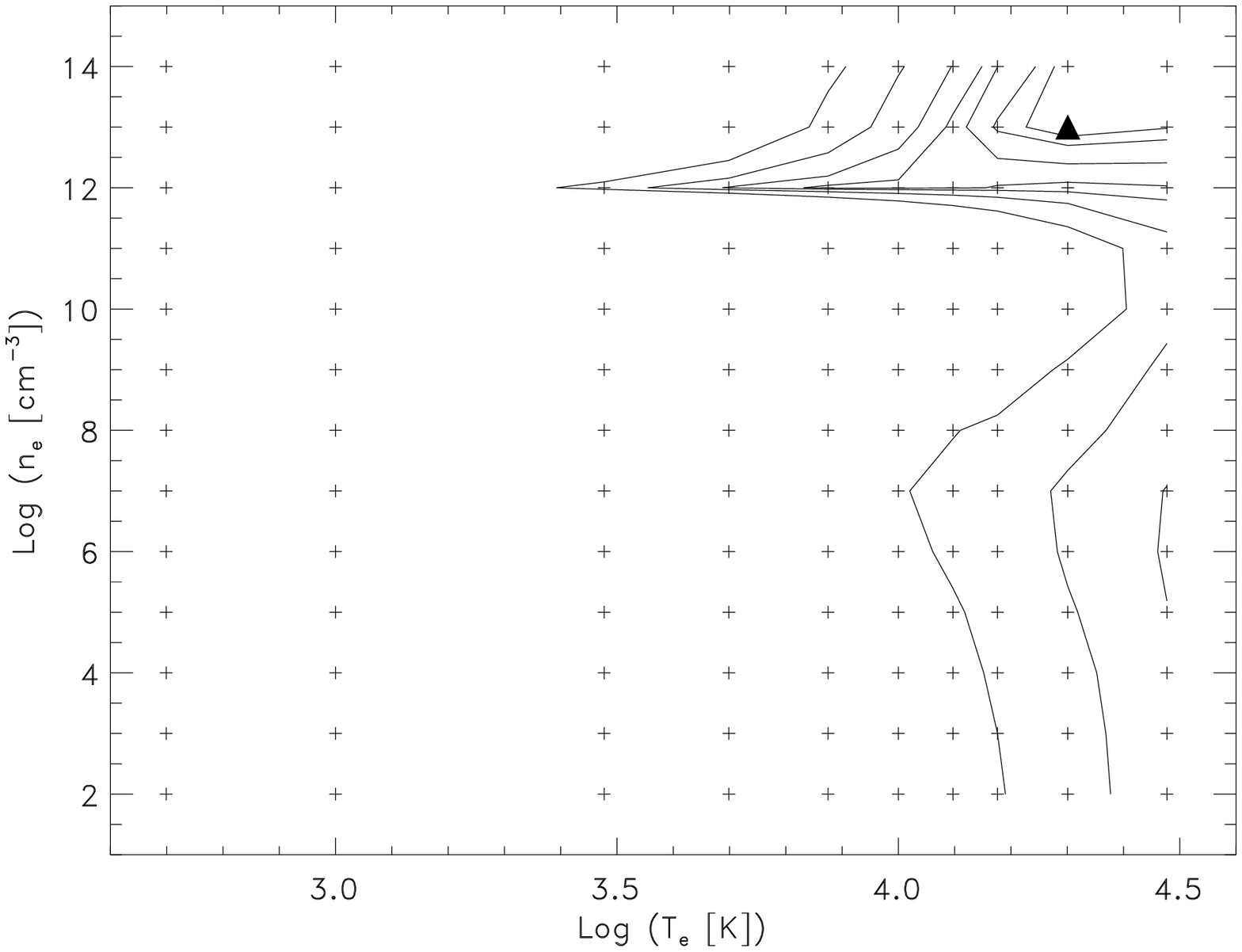}
\caption{Contours of $\chi^2$ values (relative to $\chi^2_{min}$) as a function of electron density $n_e$ and temperature $T_e$ from fits of the theoretical H line fluxes to the values derived from the continuum subtracted observations of TW Hya. Contour levels are 1.05, 1.1, 1.2, 1.3, 1.5, 2.0, and 3.0 $\chi^2_{min}$. Pluses denote the temperature and density combinations for which theoretical line ratios were computed and compared to the observations. The solid triangle denotes the best fit. }
\label{chicont}
\end{figure}

\begin{figure}
\includegraphics[width=6in, keepaspectratio=true]{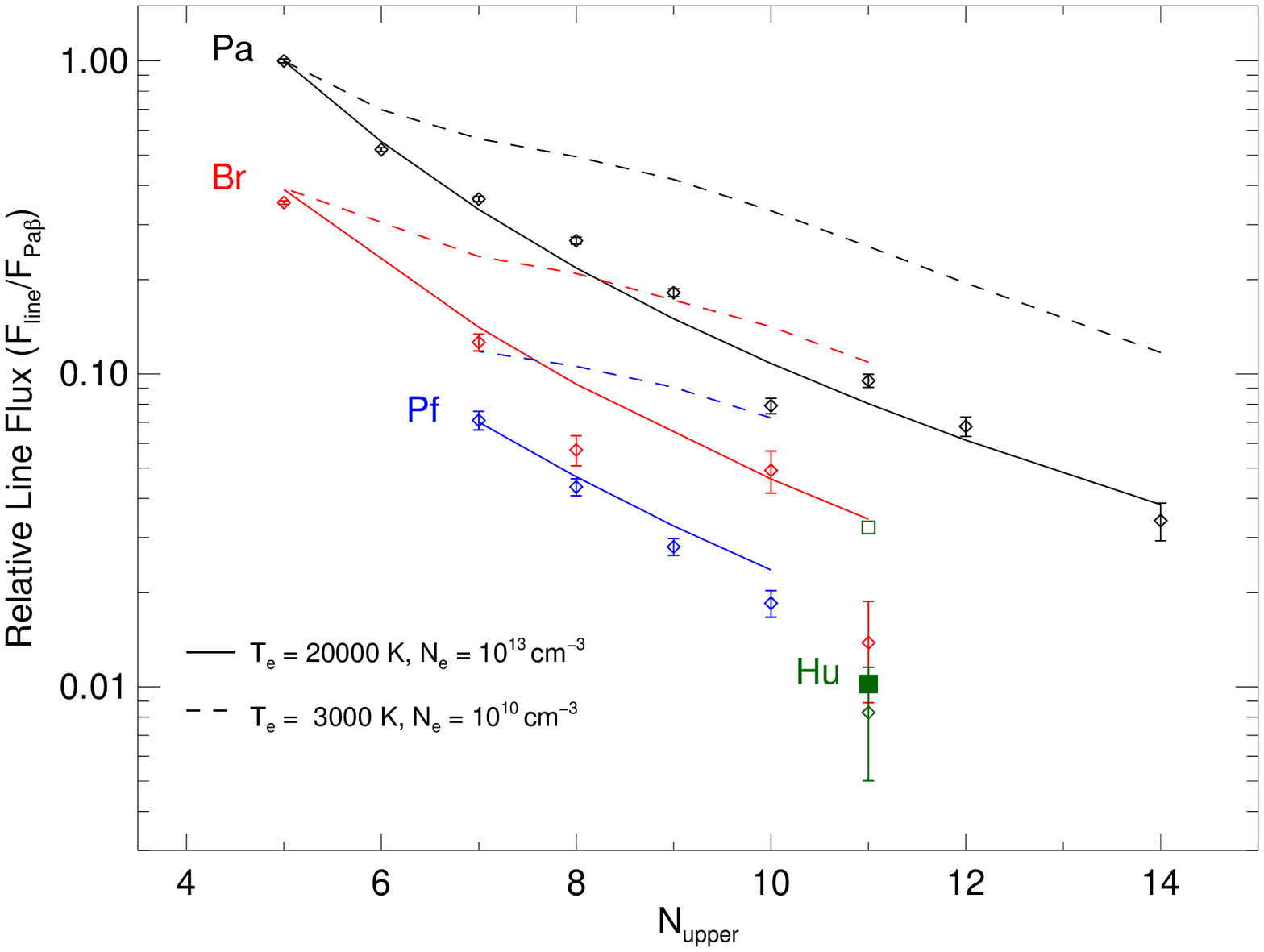}
\caption{Observed (points) and best-fitting theoretical line fluxes (solid lines, corresponding to $T_e = 2\times 10^4$ K and $n_e=10^{13}$ cm$^{-3}$) relative to Pa $\beta$. We detected only a single line of the Humphreys series (Hu 11);  the solid green square denotes the theoretical value for this line. The dotted lines and open green square denote the theoretical line fluxes for the temperature and density found by \citet{Bary08} in their study of 15 CTTS systems ($T_e = 3000$ K and $n_e=10^{10}$ cm$^{-3}$).}
\label{lineratio}
\end{figure}


\begin{figure}
\centering
\begin{tabular}{cc}

\includegraphics[width=2.75in, keepaspectratio=true]{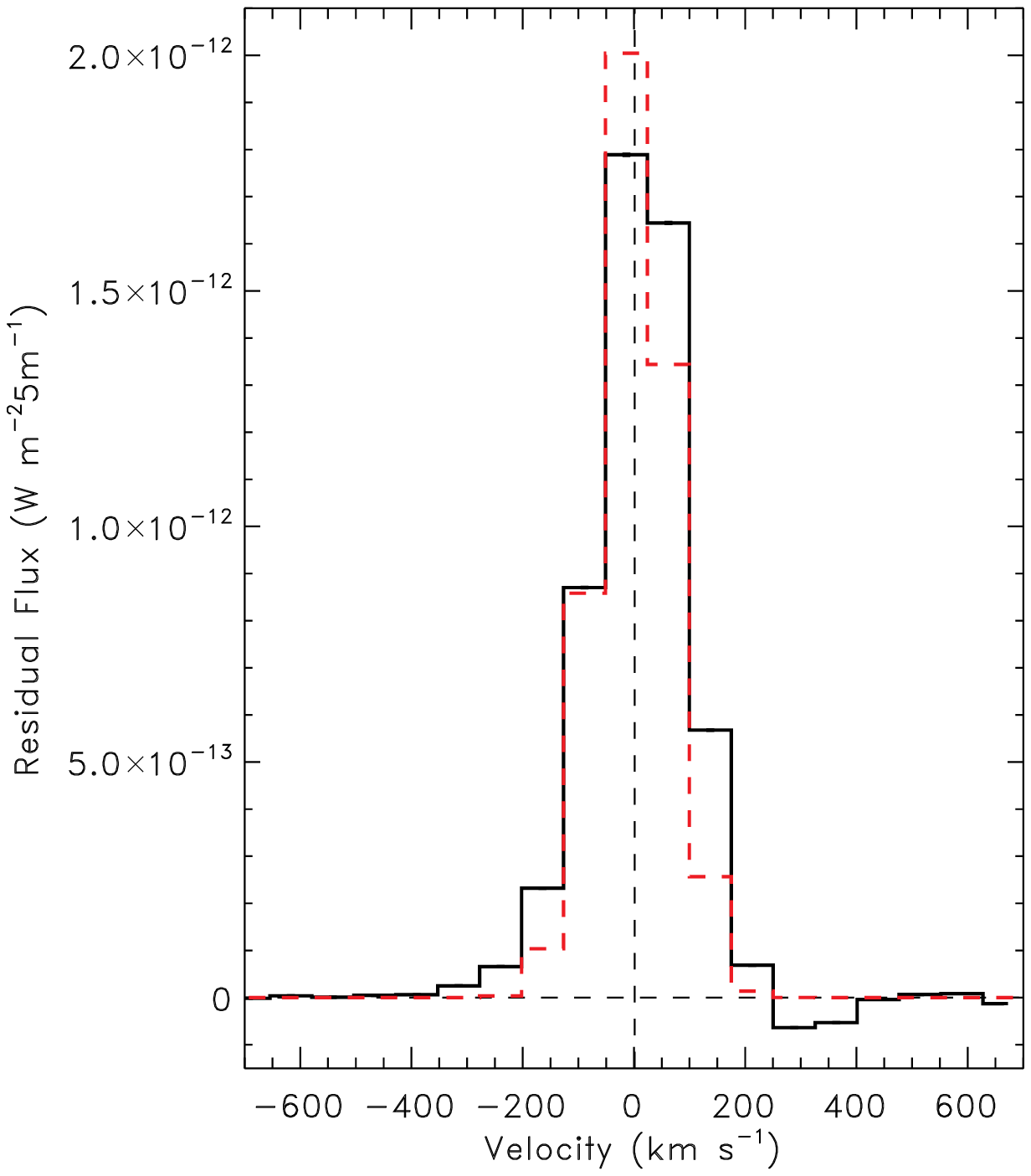}

%
\includegraphics[width=2.75in, keepaspectratio=true]{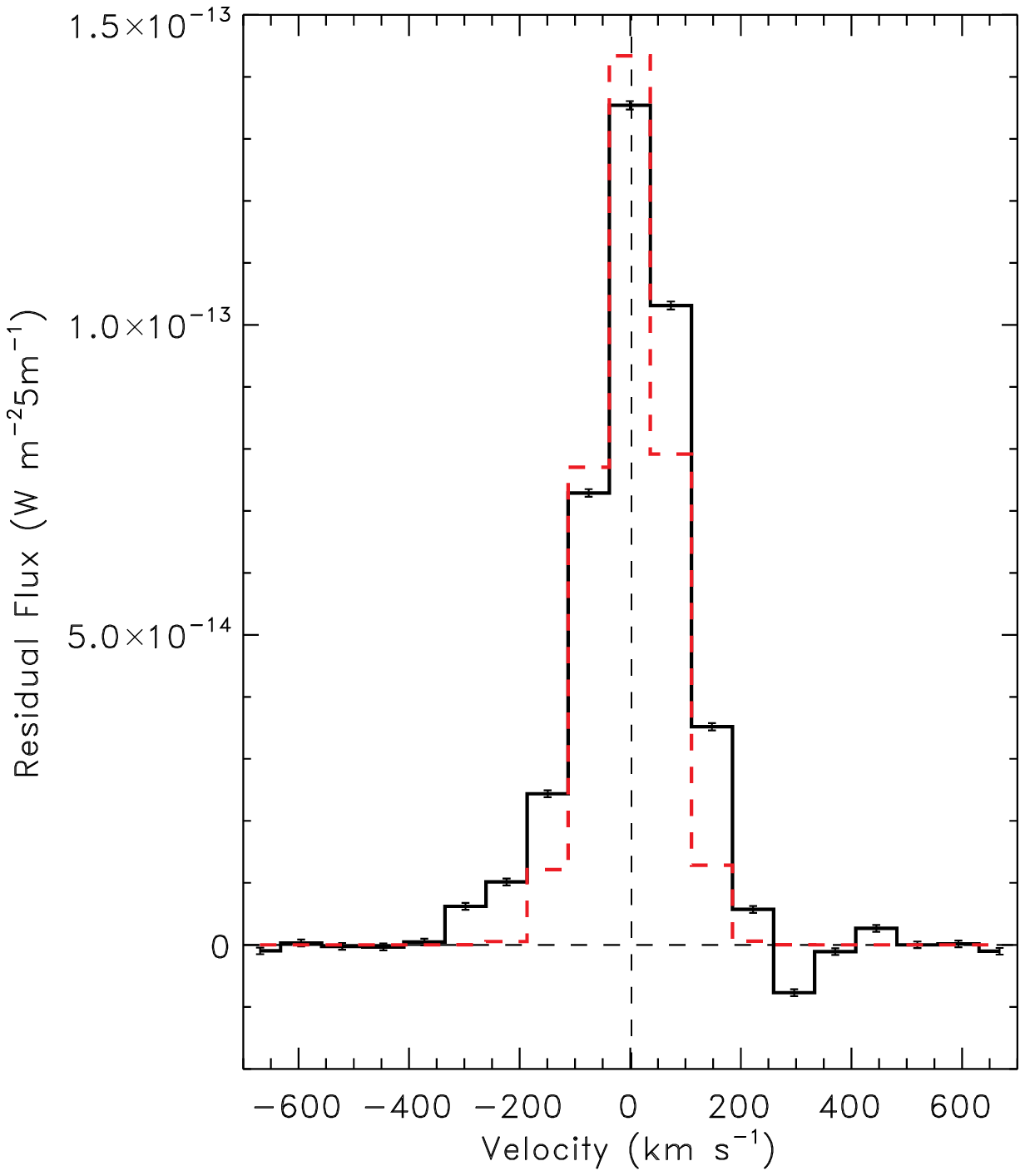}

\end{tabular}
\caption{Pa $\beta$ (left) and Br $\gamma$ (right) line profiles. Note the strong core centered at zero velocity, atop a weaker and broader component, and a weak absorption component in the red wing at 300 km s$^{-1}$. The dotted line is a Gaussian with the FWHM given by the nominal spectral resolution.}
\label{Hlineprof}
\end{figure}

\begin{figure}
\centering
\includegraphics[width=2.75in,keepaspectratio=true]{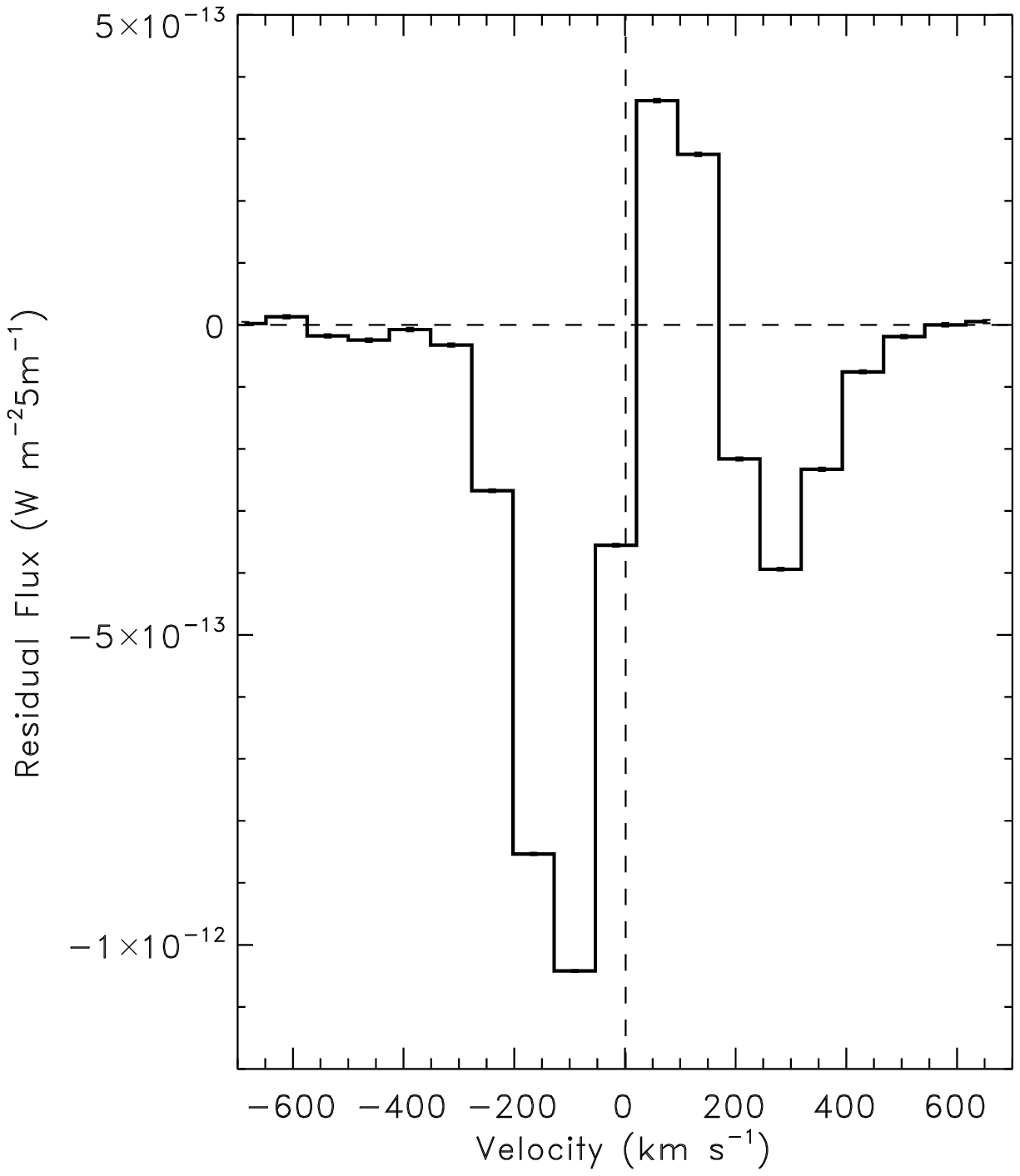}
\caption{\ion{He}{1} 1.083 $\mu$m line profile. Note the red and blue absorption components. The red absorption feature is located at the same velocity as the similar feature in the H profiles.}
\label{Helineprof}
\end{figure}


\begin{figure}
\includegraphics[height=7.5in,angle=90]{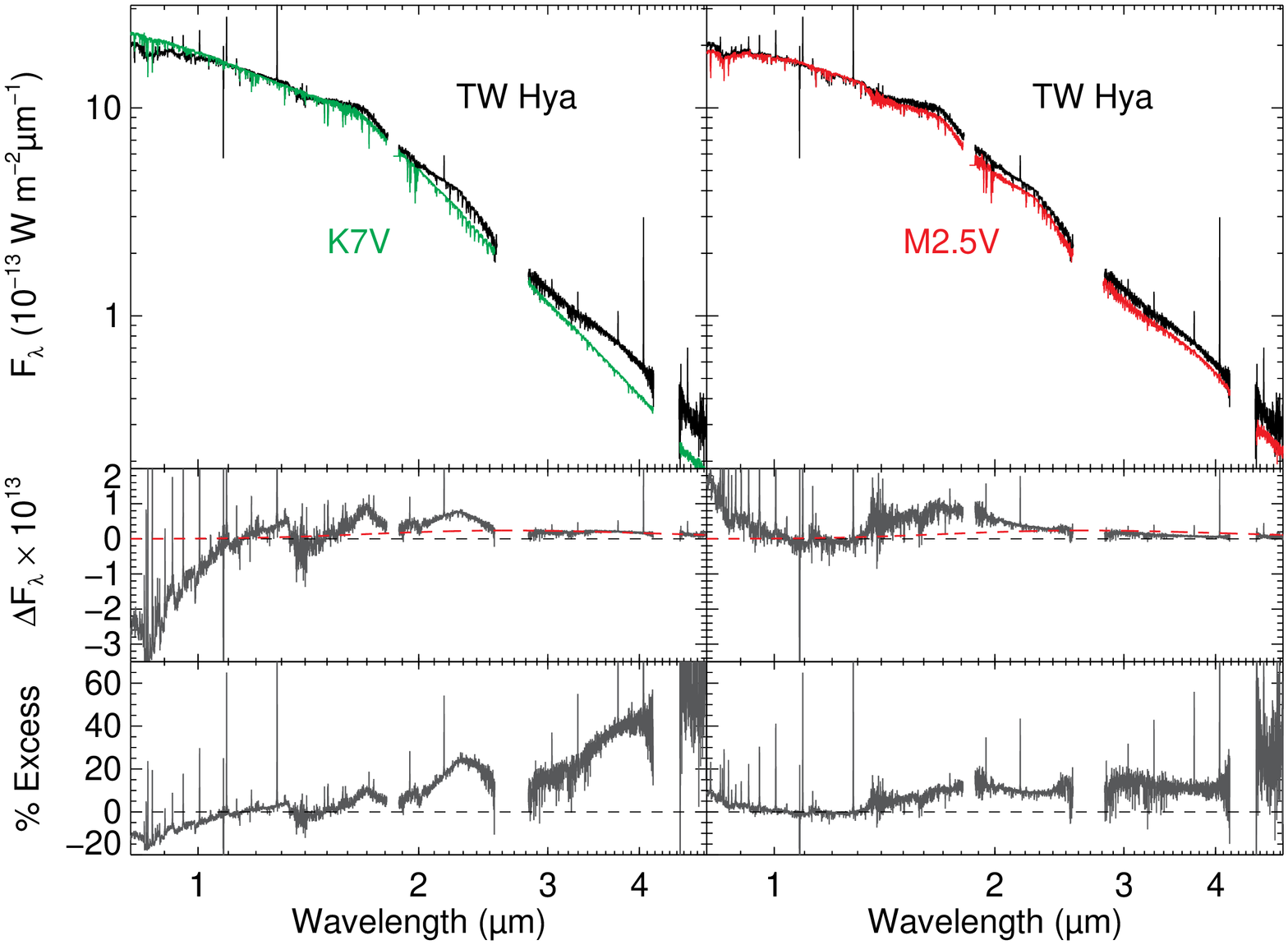}
\caption{(Top) Comparison between the spectrum of TW Hya and that of HD 237903 (K7V; left) and Gl 381 (M2.5V, right). The K7V and M2.5V spectra have been scaled to match the flux levels of TW Hya between 0.9 and 1.35 $\mu$m. (Middle) Excess flux spectra generated by subtracting the scaled K7V spectrum (left) and the scaled M2.5V spectrum (right) from the observed spectrum of TW Hya. The red dashed line denotes the excess flux predicted by the model of \citet{Eisner06}. (Bottom) The percentage excess relative to the scaled template spectra.}
\label{veilplot}
\end{figure}

\begin{figure}
\includegraphics[keepaspectratio=true,width=5in,angle=90]{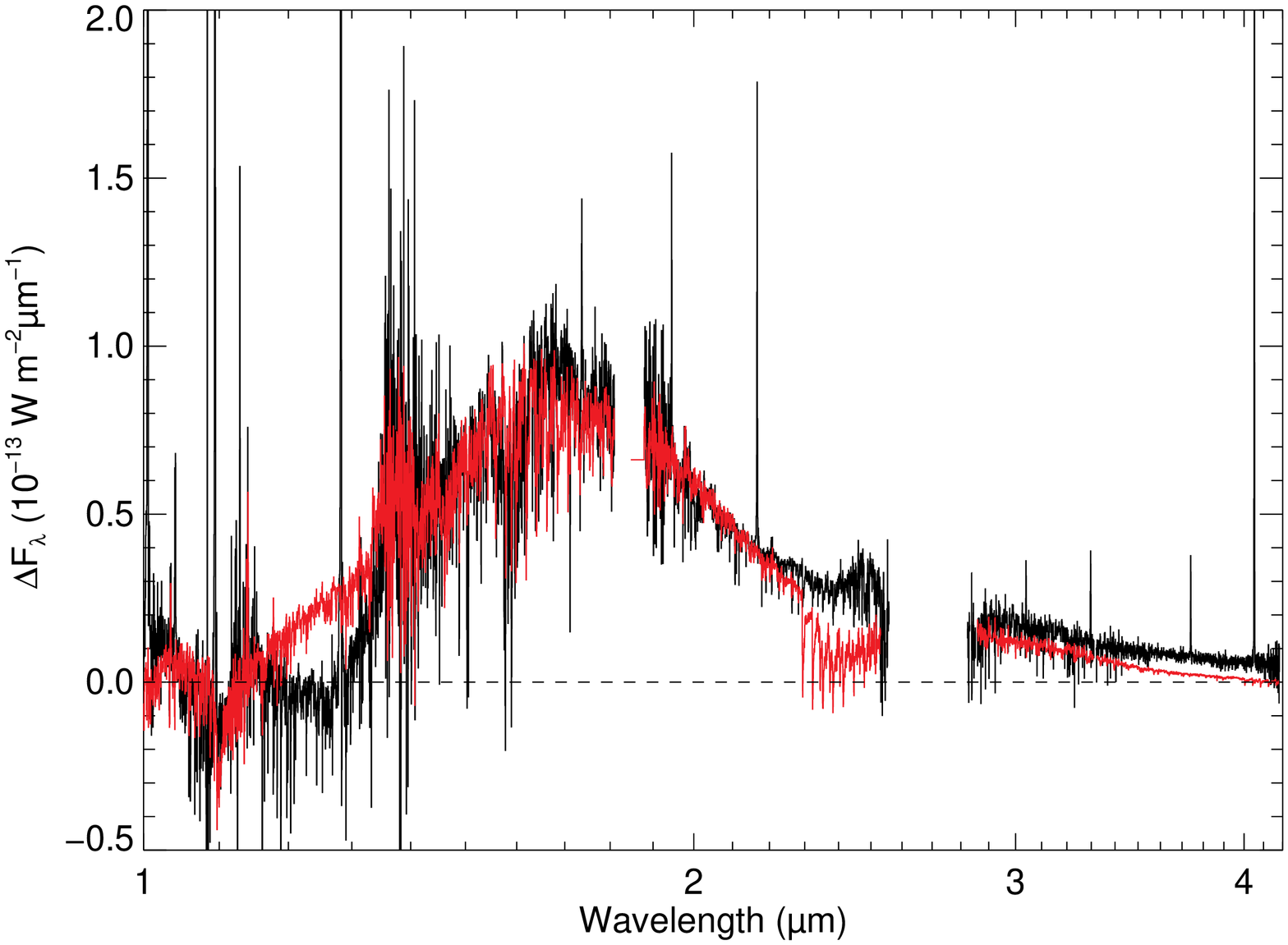}
\caption{The excess flux spectrum (black) generated by subtracting a scaled M2.5V spectrum from the spectrum of TW Hya, compared with the excess flux spectrum (red) generated by subtracting a scaled M2.5V spectrum from an M2.5III spectrum. The latter excess spectrum has been scaled to match the TW Hya excess spectrum between 1.3 and 2.2 $\mu$m. }
\label{giant_dwarf}
\end{figure}

\begin{figure}
\includegraphics[keepaspectratio=true,width=6in]{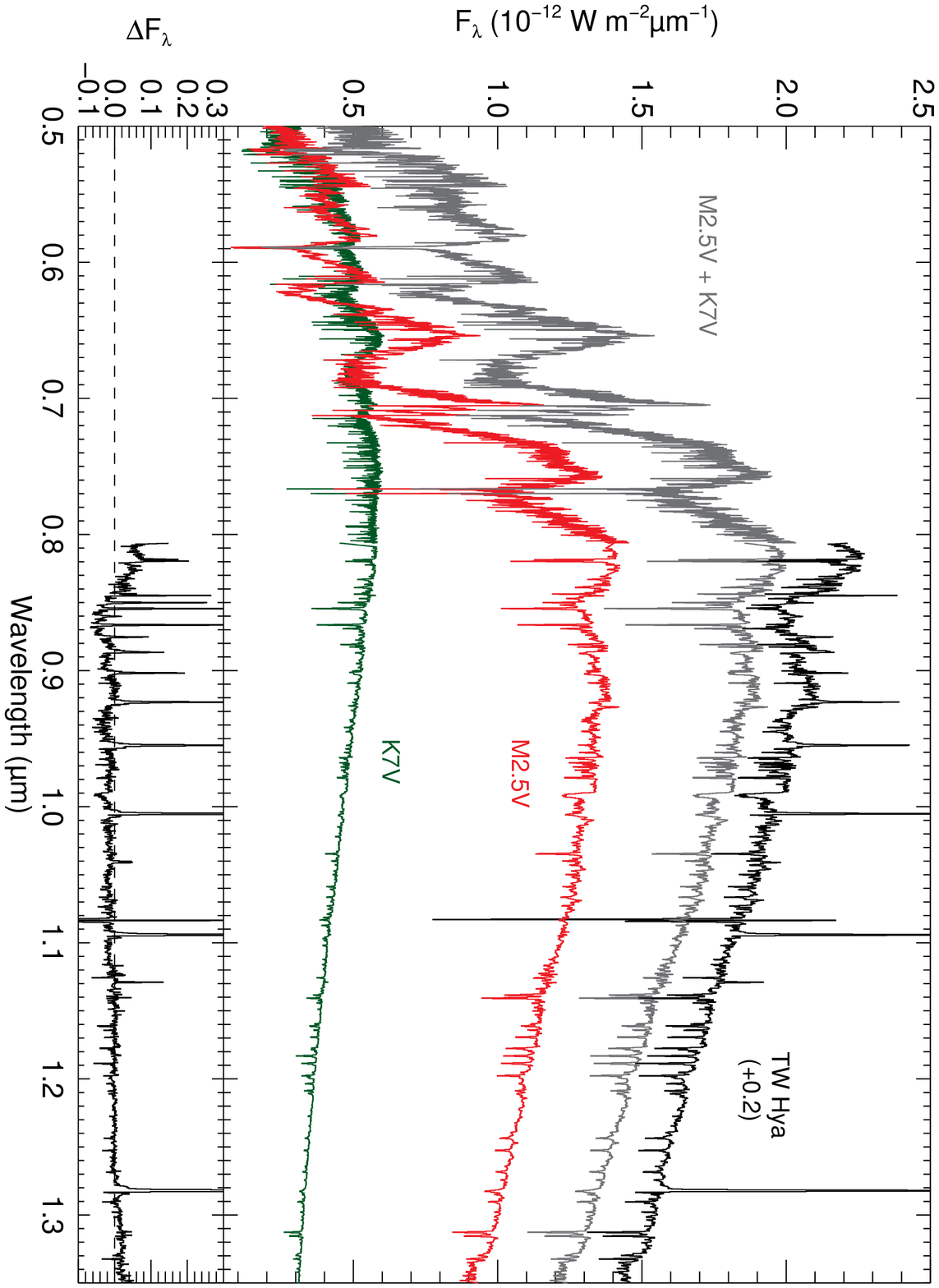}
\caption{The observed NIR spectrum of TW Hya (black line, offset by 0.2) compared to a composite spectrum (grey) generated from the sum of an M2.5V (red) and a K7V (green) spectrum.}
\label{M+K}
\end{figure}

\begin{figure}
\includegraphics[keepaspectratio=true,width=7.5in]{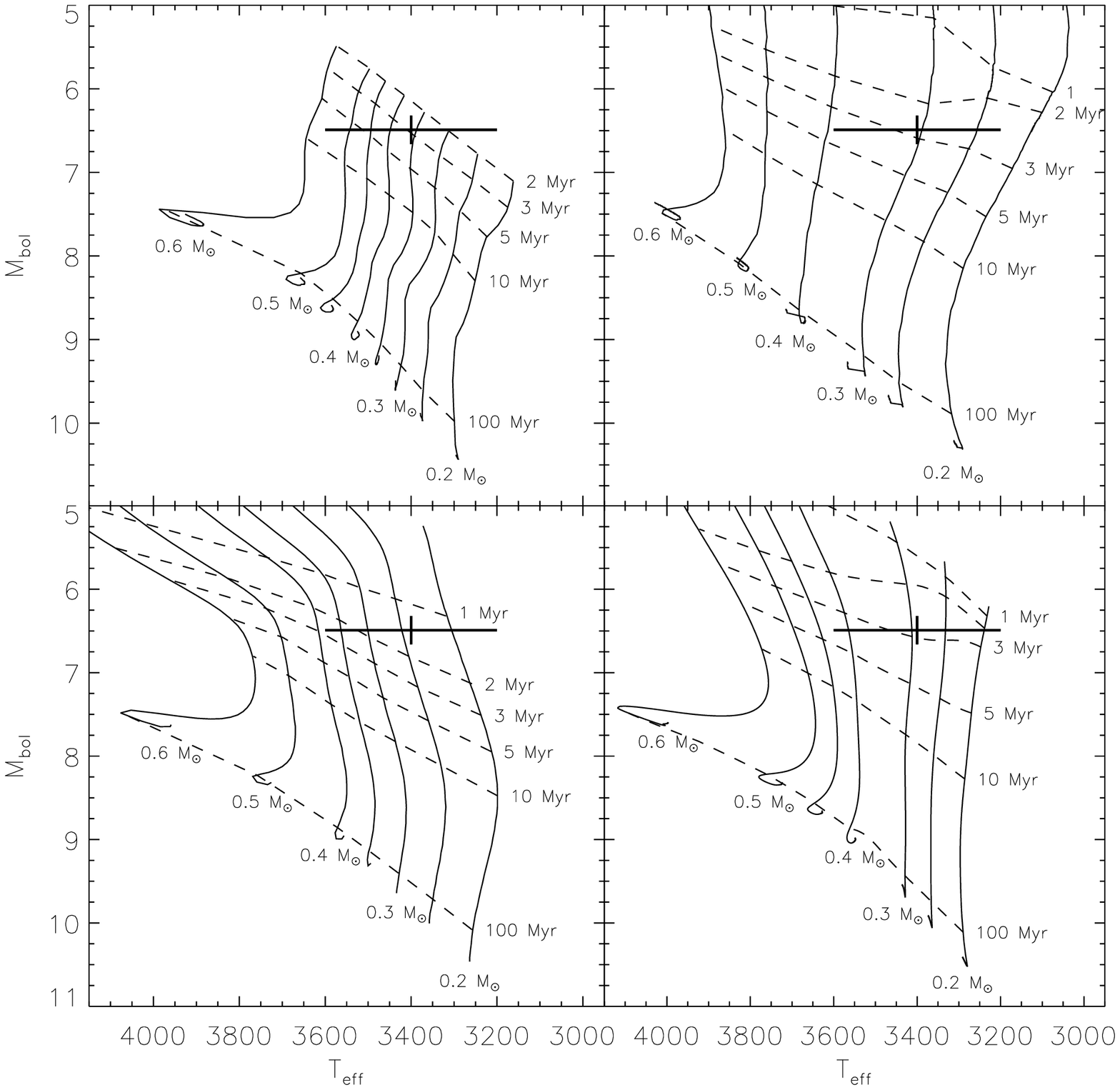}
\caption{H-R Diagram for TW Hya (thick cross). Overplotted are the theoretical tracks (solid lines) and isochrones (dashed lines) of \citet{Baraffe98} (upper left), \citet{D'Antona98} (lower left), \citet{Siess00} (upper right), and \citet{Tognelli10} (lower right).}
\label{HRD}
\end{figure}

\newpage

\begin{deluxetable}{l l c c}
\tablewidth{0pt}
\tablecaption{EWs of Selected Absorption Features in the NIR spectrum of TW Hya}
\tablehead{
\colhead{Feature} & \colhead{$\lambda$} &\colhead{EW} & \colhead{$\sigma$} \\
                                  & \colhead{$\mu$m} & \colhead{(\AA)} & \colhead{(\AA)} 
}
\startdata

\ion{Al}{1} & 1.313 & 2.38 & 0.08 \\
\ion{Al}{1} & 1.675  & 4.13 & 0.17 \\
\ion{Al}{1} & 2.110 & 0.80 & 0.08 \\
\ion{Al}{1} & 2.117 & 1.12 & 0.08 \\
\ion{K}{1} & 1.253 & 0.72 & 0.04 \\
\ion{Mg}{1} & 1.183 & 0.97 & 0.03 \\
\ion{Mg}{1} & 1.504 & 4.81 & 0.14 \\
\ion{Mg}{1} &  1.488 & 1.51 & 0.13 \\
\ion{Mg}{1} & 1.711 & 2.61 & 0.21 \\
\ion{Mg}{1} & 2.107 & 0.50 & 0.07 \\
\ion{Na}{1} & 1.140 &  2.94 & 0.34 \\
\ion{Na}{1} & 2.206 & 4.41& 0.08\\
\ion{Na}{1} & 2.336 & 2.12 & 0.16 \\
FeH & 0.99 & 3.10 & 0.08 \\
\enddata

\label{abslines}
\end{deluxetable}

\newpage

\begin{deluxetable}{l c c l}
\tabletypesize{\small}
\tablewidth{0pt}
\tablecaption{Fluxes of Emission Lines in the NIR spectrum of TW Hya}
\tablehead{
\colhead{$\lambda_{\rm obs}$\tablenotemark{a}} &\colhead{Flux} & \colhead{$\sigma$} & Identification \\
                     \colhead{($\mu$m)}   & \multicolumn{2}{c}{($10^{-17}~ {\rm W ~m^{-2}}$)} & \\ 
}
\startdata
0.8183 & 8.21    & 0.63 & \ion{Na}{1} \\
0.8195 & 11.18  & 0.71 & \ion{Na}{1} \\
0.8448 & 13.61 & 0.58 & \ion{O}{1}\\
0.8502 & 17.03 & 0.83 & \ion{Ca}{2} + Pa 16 \\
0.8545 & 19.16 & 0.70 & \ion{Ca}{2} + Pa 15 \\
0.8602 & 6.01 & 0.82 & Pa 14 \\
0.8630 & 1.85 & 0.56 & ? \\
0.8665 & 18.52 & 0.75 & \ion{Ca}{2} + Pa 13 \\
0.8753 & 12.02 & 0.84 & Pa 12 \\
0.8864 & 16.83 & 0.80 & Pa 11 \\
0.9018 & 13.99 & 0.79 & Pa 10 \\
0.9232 & 32.16 & 0.88 & Pa 9 \\
0.9549 & 47.15 & 1.09 & Pa 8 ($\epsilon$) \\
1.0051 & 64.00 & 1.00 & Pa 7 ($\delta$)\\
1.0406 & 9.01 & 1.04 & [\ion{N}{1}]?\\
1.0941 & 92.11 & 0.92 & Pa 6 ($\gamma$)\\
1.1289 & 9.99 & 0.40 & \ion{O}{1} \\
1.2821 & 176.89 & 1.79 & Pa 5 ($\beta$)\\
1.6812 & 2.45 & 0.87 & Br 11 \\
1.7368 & 8.69 & 1.33 & Br 10 \\
1.9449 & 10.11 & 1.11 & Br 8 ($\epsilon$)\\
2.1661 & 22.32 & 1.37 & Br 7 ($\gamma$)\\
3.0395 & 3.27 & 0.32 & Pf 10 ($\epsilon$) \\
3.2971 & 4.96 & 0.30 & Pf 9 ($\delta$)\\
3.7404 & 7.70 & 0.47 & Pf 8 ($\gamma$) \\
4.0523 & 62.38 & 0.55 & Br 5 ($\alpha$)\\
4.6538 & 12.57 & 0.84 & Pf 7 ($\beta$)\\
4.6727 & 1.46 & 0.58 & Hu 11 ($\epsilon$)\\

\enddata

\tablenotetext{a}{Observed vacuum wavelength}
\label{emisslines}
\end{deluxetable}

\end{document}